\documentclass[12pt, draftclsnofoot, onecolumn]{IEEEtran}
\usepackage[T1]{fontenc}
\usepackage{cite}
\ifCLASSINFOpdf
\else
\fi   
\usepackage{url}
\usepackage{amsbsy}
\usepackage{amsmath}
\usepackage{fixmath}
\usepackage{amssymb}
\usepackage{ulem}

\usepackage[noend]{algorithmic}
\usepackage{algorithm}
\newlength\myindent
\algsetup{indent=\myindent}
\algsetup{linenodelimiter=}

\usepackage{tikz}
\usetikzlibrary{arrows,decorations,backgrounds,shadows,plotmarks}

\usepackage{trfsigns}
\usepackage{amssymb}
\usepackage{amsmath}
\usepackage{adjustbox}
\usepackage{romannum}
\usepackage{booktabs}
\usepackage{caption}
\usepackage{lipsum}
\usepackage{xcolor,colortbl}
\usepackage{makecell}
\usepackage{pgfplots}
\usepackage{bm}
\newtheorem{example}{Example}
\newtheorem{assumption}{Assumption}
\usepackage{algorithm}
\usepackage{algorithmic}
\usepackage{siunitx}
\usepackage{float}
\usepackage{pgfplots}
\pgfplotsset{compat=newest}
\pgfplotsset{plot coordinates/math parser=false}
\pgfplotsset{every axis/.append style={font=\footnotesize}}
\newlength\figureheight
\newlength\figurewidth
    \definecolor{TUMBeamerYellow}    {rgb} {1.000,0.706,0.000}    
    \definecolor{TUMBeamerOrange}    {rgb} {1.000,0.502,0.000}    
    \definecolor{TUMBeamerRed}       {rgb} {0.898,0.204,0.094}    
    \definecolor{TUMBeamerDarkRed}   {rgb} {0.792,0.129,0.247}    
    \definecolor{TUMBeamerBlue}      {rgb} {0.000,0.600,1.000}    
    \definecolor{TUMBeamerLightBlue} {rgb} {0.255,0.745,1.000}    
    \definecolor{TUMBeamerGreen}     {rgb} {0.569,0.675,0.420}    
    \definecolor{TUMBeamerLightGreen}{rgb} {0.710,0.792,0.510}    

\begin{document}
\title{Learning The MMSE Channel Predictor}
\author{Nurettin~Turan and Wolfgang~Utschick\thanks{The authors are with Methods of Signal Processing, Technische Universit\"at M\"unchen, 80290 Munich, Germany (e-mail: \{nurettin.turan,utschick\}@tum.de.}
}
\markboth{}{}

\maketitle

\begin{abstract}
We present a neural network based predictor which is derived by starting from the linear minimum mean squared error (LMMSE) predictor and by further making two key assumptions. With these assumptions, we first derive a weighted sum of LMMSE predictors which is motivated by the structure of the optimal MMSE predictor. This predictor provides an initialization (weight matrices, biases and activation function) to a feed-forward neural network based predictor. With a properly learned neural network, we show that under the given channel model assumptions it is possible to easily outperform the LMMSE predictor based on the Jakes assumption of the underlying Doppler spectrum.
\end{abstract}


\IEEEpeerreviewmaketitle

\section{Introduction}
For increasing the achievable transmission rate in a wireless communication system, it is beneficial to have channel state information (CSI) at the transmitter side \cite{Zemen}. In scenarios, where the users are moving, the CSI may get outdated rapidly; thus, channel prediction plays an important role \cite{Zemen}.

After reformulating the general expression of the linear MMSE (LMMSE) predictor, a similar approach as for the learning based low complexity MMSE channel estimator \cite{Neumann} is used to obtain a learning based MMSE channel predictor. The starting point to the channel prediction problem is an underlying channel model. By reformulating the LMMSE predictor and by further making two key assumptions it is possible to derive a weighted sum of LMMSE predictors which has the structure of a feed-forward neural network. With a properly learned neural network, we show that under the given channel model assumptions it is possible to easily outperform the LMMSE predictor based on the Jakes assumption of the underlying Doppler spectrum.

\section{Channel Model}

In this Section, the fading process and thus the underlying physical wave propagation channel model is explained, which is used to evaluate the performances of the predictors derived in the following sections. 
The channel is constructed by the superposition of $P$ impinging plane-waves, where each plane-wave propagates along a specific path and is eventually received by a moving user with constant velocity $v$ ~\cite{Clarke,Zemen}. Each of these paths is mainly determined by a path-specific Doppler shift $f_p$ and path phase $\psi_p$, where the Doppler shift of path $p$ equals to:
\begin{equation}
        f_p = \dfrac{v f_c}{c} \cos{\delta_p},
\end{equation}
with $f_c$ being the carrier frequency, $c$ the speed of light and $\delta_p$ the direction of arrival (DoA) of path $p$. The maximum possible Doppler shift is defined as the Doppler bandwidth $B_D = \dfrac{v f_c}{c}$. We assume that over a block of $M+N$ symbols, which is equals to the duration of $(M+N)T_s$, the path phases and the Doppler shifts remain unchanged, where $M$ is the \textit{observation length}, $N$ is the \textit{prediction length} and $T_s$ is the \textit{symbol duration}, which is much longer than the delay spread of the channel, thus, we have a frequency-flat channel~\cite{Zemen}. We further assume that each path-phase $\psi_p$ and each path specific DoA $\delta_p$ are uniformly distributed, i.e. \cite{Zemen, Clarke}:
\begin{equation}
        \psi_p \sim \mathcal{U}[-\pi,\pi)
\end{equation}
\begin{equation}
    \delta_p \sim \mathcal{U}[-\pi,\pi).
\end{equation}
With these assumptions, the channel coefficients $h[m]$ for $m=0,\dots, M+N-1$ are constructed by ~\cite{Clarke,Zemen}:
\begin{equation}
    h[m] = \sum_{p=0}^{P-1} \dfrac{1}{\sqrt{P}} e^{j\psi_p} e^{j2\pi f_p T_s m} = \sum_{p=0}^{P-1} a_p e^{j2\pi f_p T_s m}.
    \label{channelcoeff}
\end{equation}
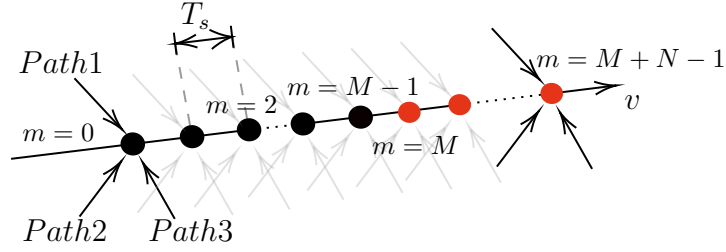
\begin{figure}
\centering
    \begin{center}
    	\end{center}

    	\tikzset{every picture/.style={line width=0.75pt}} 
    	
    	\begin{tikzpicture}[x=0.75pt,y=0.75pt,yscale=-1,xscale=1]
    	
    	\draw [color={rgb, 255:red, 155; green, 155; blue, 155 }  ,draw opacity=1 ] [dash pattern={on 4.5pt off 4.5pt}]  (89.6,23.53) -- (98.14,70.47) ;

    	\draw [color={rgb, 255:red, 155; green, 155; blue, 155 }  ,draw opacity=1 ] [dash pattern={on 4.5pt off 4.5pt}]  (118.34,20.06) -- (126.88,67) ;

    	\draw    (154.02,64.05) -- (233.08,54.33) ;

    	\draw  [dash pattern={on 0.84pt off 2.51pt}]  (233.08,54.33) -- (279.5,48.66) ;

    	\draw  [fill={rgb, 255:red, 0; green, 0; blue, 0 }  ,fill opacity=1 ] (121.24,66.28) .. controls (121.61,63.36) and (124.44,61.32) .. (127.55,61.72) .. controls (130.67,62.12) and (132.89,64.81) .. (132.52,67.73) .. controls (132.15,70.65) and (129.32,72.69) .. (126.2,72.29) .. controls (123.09,71.89) and (120.86,69.2) .. (121.24,66.28) -- cycle ;
    	\draw  [fill={rgb, 255:red, 0; green, 0; blue, 0 }  ,fill opacity=1 ] (92.49,69.75) .. controls (92.87,66.83) and (95.7,64.78) .. (98.81,65.18) .. controls (101.93,65.59) and (104.15,68.28) .. (103.78,71.2) .. controls (103.41,74.11) and (100.58,76.16) .. (97.46,75.76) .. controls (94.35,75.36) and (92.12,72.66) .. (92.49,69.75) -- cycle ;
    	\draw  [fill={rgb, 255:red, 0; green, 0; blue, 0 }  ,fill opacity=1 ] (62.57,73.51) .. controls (62.89,70.59) and (65.68,68.5) .. (68.8,68.85) .. controls (71.92,69.2) and (74.19,71.85) .. (73.88,74.78) .. controls (73.56,77.7) and (70.77,79.79) .. (67.65,79.44) .. controls (64.53,79.09) and (62.25,76.44) .. (62.57,73.51) -- cycle ;
    	\draw  [fill={rgb, 255:red, 0; green, 0; blue, 0 }  ,fill opacity=1 ] (148.38,63.32) .. controls (148.75,60.4) and (151.58,58.36) .. (154.7,58.76) .. controls (157.82,59.16) and (160.04,61.85) .. (159.67,64.77) .. controls (159.29,67.69) and (156.47,69.73) .. (153.35,69.33) .. controls (150.23,68.93) and (148.01,66.24) .. (148.38,63.32) -- cycle ;
    	\draw  [fill={rgb, 255:red, 11; green, 0; blue, 0 }  ,fill opacity=1 ] (177.5,59.9) .. controls (177.87,56.98) and (180.7,54.94) .. (183.82,55.34) .. controls (186.93,55.74) and (189.16,58.43) .. (188.78,61.35) .. controls (188.41,64.27) and (185.58,66.31) .. (182.47,65.91) .. controls (179.35,65.51) and (177.13,62.82) .. (177.5,59.9) -- cycle ;
    	\draw    (279.5,48.66) -- (310.16,44.86) ;
    	\draw [shift={(312.14,44.62)}, rotate = 532.94] [color={rgb, 255:red, 0; green, 0; blue, 0 }  ][line width=0.75]    (10.93,-3.29) .. controls (6.95,-1.4) and (3.31,-0.3) .. (0,0) .. controls (3.31,0.3) and (6.95,1.4) .. (10.93,3.29)   ;
    	
    	\draw    (38.8,41) -- (61.65,65.57) ;
    	\draw [shift={(63.01,67.03)}, rotate = 227.07999999999998] [color={rgb, 255:red, 0; green, 0; blue, 0 }  ][line width=0.75]    (10.93,-3.29) .. controls (6.95,-1.4) and (3.31,-0.3) .. (0,0) .. controls (3.31,0.3) and (6.95,1.4) .. (10.93,3.29)   ;
    	
    	\draw    (41.37,110.8) -- (61.34,84.32) ;
    	\draw [shift={(62.54,82.73)}, rotate = 487.02] [color={rgb, 255:red, 0; green, 0; blue, 0 }  ][line width=0.75]    (10.93,-3.29) .. controls (6.95,-1.4) and (3.31,-0.3) .. (0,0) .. controls (3.31,0.3) and (6.95,1.4) .. (10.93,3.29)   ;
    	
    	\draw    (89.4,111.97) -- (74.15,85.12) ;
    	\draw [shift={(73.16,83.38)}, rotate = 420.40999999999997] [color={rgb, 255:red, 0; green, 0; blue, 0 }  ][line width=0.75]    (10.93,-3.29) .. controls (6.95,-1.4) and (3.31,-0.3) .. (0,0) .. controls (3.31,0.3) and (6.95,1.4) .. (10.93,3.29)   ;

    	\draw [color={rgb, 255:red, 155; green, 155; blue, 155 }  ,draw opacity=0.3 ]   (68.12,39.04) -- (90.96,63.6) ;
    	\draw [shift={(92.33,65.07)}, rotate = 227.07999999999998] [color={rgb, 255:red, 155; green, 155; blue, 155 }  ,draw opacity=0.3 ][line width=0.75]    (10.93,-3.29) .. controls (6.95,-1.4) and (3.31,-0.3) .. (0,0) .. controls (3.31,0.3) and (6.95,1.4) .. (10.93,3.29)   ;
    	
    	\draw [color={rgb, 255:red, 155; green, 155; blue, 155 }  ,draw opacity=0.3 ]   (70.68,108.84) -- (90.65,82.36) ;
    	\draw [shift={(91.85,80.76)}, rotate = 487.02] [color={rgb, 255:red, 155; green, 155; blue, 155 }  ,draw opacity=0.3 ][line width=0.75]    (10.93,-3.29) .. controls (6.95,-1.4) and (3.31,-0.3) .. (0,0) .. controls (3.31,0.3) and (6.95,1.4) .. (10.93,3.29)   ;
    	
    	\draw [color={rgb, 255:red, 155; green, 155; blue, 155 }  ,draw opacity=0.3 ]   (118.71,110) -- (103.46,83.15) ;
    	\draw [shift={(102.48,81.41)}, rotate = 420.40999999999997] [color={rgb, 255:red, 155; green, 155; blue, 155 }  ,draw opacity=0.3 ][line width=0.75]    (10.93,-3.29) .. controls (6.95,-1.4) and (3.31,-0.3) .. (0,0) .. controls (3.31,0.3) and (6.95,1.4) .. (10.93,3.29)   ;

    	\draw    (249.29,15.77) -- (272.13,40.33) ;
    	\draw [shift={(273.5,41.8)}, rotate = 227.07999999999998] [color={rgb, 255:red, 0; green, 0; blue, 0 }  ][line width=0.75]    (10.93,-3.29) .. controls (6.95,-1.4) and (3.31,-0.3) .. (0,0) .. controls (3.31,0.3) and (6.95,1.4) .. (10.93,3.29)   ;
    	
    	\draw    (251.85,85.57) -- (271.82,59.09) ;
    	\draw [shift={(273.02,57.5)}, rotate = 487.02] [color={rgb, 255:red, 0; green, 0; blue, 0 }  ][line width=0.75]    (10.93,-3.29) .. controls (6.95,-1.4) and (3.31,-0.3) .. (0,0) .. controls (3.31,0.3) and (6.95,1.4) .. (10.93,3.29)   ;
    	
    	\draw    (299.88,86.73) -- (284.64,59.88) ;
    	\draw [shift={(283.65,58.14)}, rotate = 420.40999999999997] [color={rgb, 255:red, 0; green, 0; blue, 0 }  ][line width=0.75]    (10.93,-3.29) .. controls (6.95,-1.4) and (3.31,-0.3) .. (0,0) .. controls (3.31,0.3) and (6.95,1.4) .. (10.93,3.29)   ;

    	\draw  [dash pattern={on 0.84pt off 2.51pt}]  (126.88,67) -- (154.02,64.05) ;

    	\draw    (6.85,81.66) -- (126.88,67) ;

    	\draw  [draw opacity=0][fill={rgb, 255:red, 229; green, 52; blue, 24 }  ,fill opacity=1 ] (202.05,57.45) .. controls (202.42,54.54) and (205.25,52.49) .. (208.37,52.89) .. controls (211.48,53.29) and (213.71,55.99) .. (213.34,58.9) .. controls (212.96,61.82) and (210.13,63.87) .. (207.02,63.47) .. controls (203.9,63.06) and (201.68,60.37) .. (202.05,57.45) -- cycle ;
    	\draw    (89.6,23.53) -- (118.34,20.06) ;
    	\draw [shift={(118.34,20.06)}, rotate = 533.12] [color={rgb, 255:red, 0; green, 0; blue, 0 }  ][line width=0.75]    (0,5.59) -- (0,-5.59)(10.93,-4.9) .. controls (6.95,-2.3) and (3.31,-0.67) .. (0,0) .. controls (3.31,0.67) and (6.95,2.3) .. (10.93,4.9)   ;
    	\draw [shift={(89.6,23.53)}, rotate = 353.12] [color={rgb, 255:red, 0; green, 0; blue, 0 }  ][line width=0.75]    (0,5.59) -- (0,-5.59)(10.93,-3.29) .. controls (6.95,-1.4) and (3.31,-0.3) .. (0,0) .. controls (3.31,0.3) and (6.95,1.4) .. (10.93,3.29)   ;
    	\draw [color={rgb, 255:red, 155; green, 155; blue, 155 }  ,draw opacity=0.3 ]   (96.77,36.27) -- (119.62,60.84) ;
    	\draw [shift={(120.98,62.3)}, rotate = 227.07999999999998] [color={rgb, 255:red, 155; green, 155; blue, 155 }  ,draw opacity=0.3 ][line width=0.75]    (10.93,-3.29) .. controls (6.95,-1.4) and (3.31,-0.3) .. (0,0) .. controls (3.31,0.3) and (6.95,1.4) .. (10.93,3.29)   ;
    	
    	\draw [color={rgb, 255:red, 155; green, 155; blue, 155 }  ,draw opacity=0.3 ]   (99.33,106.08) -- (119.3,79.6) ;
    	\draw [shift={(120.5,78)}, rotate = 487.02] [color={rgb, 255:red, 155; green, 155; blue, 155 }  ,draw opacity=0.3 ][line width=0.75]    (10.93,-3.29) .. controls (6.95,-1.4) and (3.31,-0.3) .. (0,0) .. controls (3.31,0.3) and (6.95,1.4) .. (10.93,3.29)   ;
    	
    	\draw [color={rgb, 255:red, 155; green, 155; blue, 155 }  ,draw opacity=0.3 ]   (147.36,107.24) -- (132.12,80.39) ;
    	\draw [shift={(131.13,78.65)}, rotate = 420.40999999999997] [color={rgb, 255:red, 155; green, 155; blue, 155 }  ,draw opacity=0.3 ][line width=0.75]    (10.93,-3.29) .. controls (6.95,-1.4) and (3.31,-0.3) .. (0,0) .. controls (3.31,0.3) and (6.95,1.4) .. (10.93,3.29)   ;

    	\draw [color={rgb, 255:red, 155; green, 155; blue, 155 }  ,draw opacity=0.3 ]   (123.91,31.31) -- (146.76,55.88) ;
    	\draw [shift={(148.12,57.35)}, rotate = 227.07999999999998] [color={rgb, 255:red, 155; green, 155; blue, 155 }  ,draw opacity=0.3 ][line width=0.75]    (10.93,-3.29) .. controls (6.95,-1.4) and (3.31,-0.3) .. (0,0) .. controls (3.31,0.3) and (6.95,1.4) .. (10.93,3.29)   ;
    	
    	\draw [color={rgb, 255:red, 155; green, 155; blue, 155 }  ,draw opacity=0.3 ]   (126.47,101.12) -- (146.44,74.64) ;
    	\draw [shift={(147.65,73.04)}, rotate = 487.02] [color={rgb, 255:red, 155; green, 155; blue, 155 }  ,draw opacity=0.3 ][line width=0.75]    (10.93,-3.29) .. controls (6.95,-1.4) and (3.31,-0.3) .. (0,0) .. controls (3.31,0.3) and (6.95,1.4) .. (10.93,3.29)   ;
    	
    	\draw [color={rgb, 255:red, 155; green, 155; blue, 155 }  ,draw opacity=0.3 ]   (174.51,102.28) -- (159.26,75.43) ;
    	\draw [shift={(158.27,73.69)}, rotate = 420.40999999999997] [color={rgb, 255:red, 155; green, 155; blue, 155 }  ,draw opacity=0.3 ][line width=0.75]    (10.93,-3.29) .. controls (6.95,-1.4) and (3.31,-0.3) .. (0,0) .. controls (3.31,0.3) and (6.95,1.4) .. (10.93,3.29)   ;

    	\draw [color={rgb, 255:red, 155; green, 155; blue, 155 }  ,draw opacity=0.3 ]   (153.68,27.83) -- (176.52,52.4) ;
    	\draw [shift={(177.89,53.86)}, rotate = 227.07999999999998] [color={rgb, 255:red, 155; green, 155; blue, 155 }  ,draw opacity=0.3 ][line width=0.75]    (10.93,-3.29) .. controls (6.95,-1.4) and (3.31,-0.3) .. (0,0) .. controls (3.31,0.3) and (6.95,1.4) .. (10.93,3.29)   ;
    	
    	\draw [color={rgb, 255:red, 155; green, 155; blue, 155 }  ,draw opacity=0.3 ]   (156.24,97.64) -- (176.21,71.16) ;
    	\draw [shift={(177.41,69.56)}, rotate = 487.02] [color={rgb, 255:red, 155; green, 155; blue, 155 }  ,draw opacity=0.3 ][line width=0.75]    (10.93,-3.29) .. controls (6.95,-1.4) and (3.31,-0.3) .. (0,0) .. controls (3.31,0.3) and (6.95,1.4) .. (10.93,3.29)   ;
    	
    	\draw [color={rgb, 255:red, 155; green, 155; blue, 155 }  ,draw opacity=0.3 ]   (204.27,98.8) -- (189.02,71.95) ;
    	\draw [shift={(188.04,70.21)}, rotate = 420.40999999999997] [color={rgb, 255:red, 155; green, 155; blue, 155 }  ,draw opacity=0.3 ][line width=0.75]    (10.93,-3.29) .. controls (6.95,-1.4) and (3.31,-0.3) .. (0,0) .. controls (3.31,0.3) and (6.95,1.4) .. (10.93,3.29)   ;

    	\draw [color={rgb, 255:red, 155; green, 155; blue, 155 }  ,draw opacity=0.3 ]   (178.38,25.99) -- (201.23,50.56) ;
    	\draw [shift={(202.59,52.02)}, rotate = 227.07999999999998] [color={rgb, 255:red, 155; green, 155; blue, 155 }  ,draw opacity=0.3 ][line width=0.75]    (10.93,-3.29) .. controls (6.95,-1.4) and (3.31,-0.3) .. (0,0) .. controls (3.31,0.3) and (6.95,1.4) .. (10.93,3.29)   ;
    	
    	\draw [color={rgb, 255:red, 155; green, 155; blue, 155 }  ,draw opacity=0.3 ]   (180.95,95.8) -- (200.92,69.32) ;
    	\draw [shift={(202.12,67.72)}, rotate = 487.02] [color={rgb, 255:red, 155; green, 155; blue, 155 }  ,draw opacity=0.3 ][line width=0.75]    (10.93,-3.29) .. controls (6.95,-1.4) and (3.31,-0.3) .. (0,0) .. controls (3.31,0.3) and (6.95,1.4) .. (10.93,3.29)   ;
    	
    	\draw [color={rgb, 255:red, 155; green, 155; blue, 155 }  ,draw opacity=0.3 ]   (228.98,96.96) -- (213.73,70.11) ;
    	\draw [shift={(212.74,68.37)}, rotate = 420.40999999999997] [color={rgb, 255:red, 155; green, 155; blue, 155 }  ,draw opacity=0.3 ][line width=0.75]    (10.93,-3.29) .. controls (6.95,-1.4) and (3.31,-0.3) .. (0,0) .. controls (3.31,0.3) and (6.95,1.4) .. (10.93,3.29)   ;

    	\draw [color={rgb, 255:red, 155; green, 155; blue, 155 }  ,draw opacity=0.3 ]   (203.27,22.74) -- (226.12,47.31) ;
    	\draw [shift={(227.48,48.78)}, rotate = 227.07999999999998] [color={rgb, 255:red, 155; green, 155; blue, 155 }  ,draw opacity=0.3 ][line width=0.75]    (10.93,-3.29) .. controls (6.95,-1.4) and (3.31,-0.3) .. (0,0) .. controls (3.31,0.3) and (6.95,1.4) .. (10.93,3.29)   ;
    	
    	\draw [color={rgb, 255:red, 155; green, 155; blue, 155 }  ,draw opacity=0.3 ]   (205.83,92.55) -- (225.8,66.07) ;
    	\draw [shift={(227.01,64.47)}, rotate = 487.02] [color={rgb, 255:red, 155; green, 155; blue, 155 }  ,draw opacity=0.3 ][line width=0.75]    (10.93,-3.29) .. controls (6.95,-1.4) and (3.31,-0.3) .. (0,0) .. controls (3.31,0.3) and (6.95,1.4) .. (10.93,3.29)   ;
    	
    	\draw [color={rgb, 255:red, 155; green, 155; blue, 155 }  ,draw opacity=0.3 ]   (253.87,93.71) -- (238.62,66.86) ;
    	\draw [shift={(237.63,65.12)}, rotate = 420.40999999999997] [color={rgb, 255:red, 155; green, 155; blue, 155 }  ,draw opacity=0.3 ][line width=0.75]    (10.93,-3.29) .. controls (6.95,-1.4) and (3.31,-0.3) .. (0,0) .. controls (3.31,0.3) and (6.95,1.4) .. (10.93,3.29)   ;

    	\draw  [draw opacity=0][fill={rgb, 255:red, 229; green, 52; blue, 24 }  ,fill opacity=1 ] (227.44,53.61) .. controls (227.81,50.69) and (230.64,48.64) .. (233.76,49.05) .. controls (236.87,49.45) and (239.1,52.14) .. (238.73,55.06) .. controls (238.35,57.98) and (235.53,60.02) .. (232.41,59.62) .. controls (229.29,59.22) and (227.07,56.53) .. (227.44,53.61) -- cycle ;
    	\draw  [draw opacity=0][fill={rgb, 255:red, 229; green, 52; blue, 24 }  ,fill opacity=1 ] (273.86,47.93) .. controls (274.23,45.01) and (277.06,42.97) .. (280.18,43.37) .. controls (283.3,43.77) and (285.52,46.46) .. (285.15,49.38) .. controls (284.78,52.3) and (281.95,54.35) .. (278.83,53.94) .. controls (275.71,53.54) and (273.49,50.85) .. (273.86,47.93) -- cycle ;
    	
    	\draw (320.27,52.82) node  [align=left] {$v$};
    	\draw (30.65,68.45) node  [align=left] {\small{$m=0$}};
    	\draw (179.43,45.56) node  [align=left] {\small{$m=M-1$}};
    	\draw (319.52,31.23) node  [align=left] {\small{$m=M+N-1$}};
    	\draw (123.06,54.31) node  [align=left] {\small{$m=2$}};
    	\draw (210.96,74.42) node  [align=left] {\small{$m=M$}};
    	\draw (31.98,32.73) node  [align=left] {$Path 1$};
    	\draw (33.86,117.79) node  [align=left] {$Path 2$};
    	\draw (97.42,117.79) node  [align=left] {$Path 3$};
    	\draw (100.00,9.5) node  [align=left] {$T_s$};

    	\end{tikzpicture}
    	\begin{center}
    		
    \end{center}
    \caption{Fading process channel construction example with $P=3$}
    \label{Fig1}
\end{figure}\newline
Drawing the limit of $P\to\infty$ the channel coefficiencts follow a Gaussian distribution based on the central limit theorem. However, it is important to note that for a low number of paths the obtained channel coefficients are distributed non-Gaussian.

An example of the fading process with three propagation paths is depicted in Figure \ref{Fig1}, where the black dots represent channel coefficients in the \textit{observation interval} $\mathcal{I}_M = \{0,1,\dots, M-1\}$ and the red dots represent channel coefficients in the \textit{prediction interval} $\mathcal{I}_N = \{M,M+1,\dots, M+N-1\}$. The goal is to make use of the correlations between the channel coefficients, in order to predict any desired channel coefficient in the prediction interval $\mathcal{I}_N$, from noisy observations of the channel coefficients of the observation interval $\mathcal{I}_M$. In this way, the channel model is a time-variant block-fading model. Accordingly, the zero mean and unit variance process is wide-sense stationary over a block, which consists of the union of the observation interval $\mathcal{I}_M$ and the prediction interval $\mathcal{I}_N$ \cite{Zemen, Zemen3}.

\subsection{Covariance Function}

In this Section, the \textit{covariance function} $R_h[k]=E[h[m]h^*[m+k]]$ of the wide-sense stationary process is described. The power spectral density (PSD) of the previously described model is given by the weighted sum of Dirac pulses at the Doppler shift frequencies $f_p$~\cite{Zemen, Goldsmith}:
\begin{equation}
S_h(f) = \sum_{p=0}^{P-1} |a_p|^2 \delta(f-f_p).
\end{equation}
Consequently, the discrete covariance function $R_h[k]$ is obtained by sampling the inverse Fourier transform of the PSD \cite{Goldsmith, Bracewell2000}, viz:
\begin{equation}
R_h[k] = \sum_{p=0}^{P-1} |a_p|^2 e^{j 2 \pi f_p T_s k},
\label{coorelationFunc_finiteP}
\end{equation}
at $k = 0, 1, \dots, M+N-1$.With the specific assumptions of the channel model in \eqref{channelcoeff} and if there are infinitely many paths, i.e., $P\to\infty$, the limit of the discrete covariance functions $R_h[k]$ is equal to~\cite{Zemen,Goldsmith,Jakes}:
\begin{eqnarray}
    R_h[k]
      &=&   J_0(2\pi\dfrac{v}{c}f_c T_s k) \\ 
      &=& J_0(2\pi B_D T_s k),
      \label{coorelationFunc_infP}
\end{eqnarray}
\text{with }$k = 0, 1, \dots, M+N-1$. By collecting the channel coefficients $h[m]$ of the observation interval $\mathcal{I}_M$ in a vector $\mathbf{h}$:
\begin{equation}
    \mathbf{h} = [h[M-1], h[M-2], \dots, h[1], h[0]]^{T},
    \label{hObsInt}
\end{equation}
we obtain the covariance matrix $\mathbf{\Sigma}_\mathbf{h}$ ~\cite{Zemen}:
\begin{equation}
    \small{
    \mathbf{\Sigma}_\mathbf{h} =
\begin{bmatrix}
R_h[0]& R_h[1]   & R_h[2] & \dots  & \dots & R_h[M-2] & R_h[M-1] \\  
R_h^*[1] & R_h[0]   & R_h[1] & R_h[2] & \dots & R_h[M-3] & R_h[M-2] \\
R_h^*[2] & R^*_h[1] & R_h[0] & R_h[1] & \dots & R_h[M-4] & R_h[M-3] \\
\vdots   && \ddots &\ddots&&& \vdots \\
R_h^*[M-2] & R^*_h[M-3] &  &  & \dots & R_h[0] & R_h[1] \\
R_h^*[M-1] & R^*_h[M-2] &  &  & \dots & R^*_h[1] & R_h[0] \\
\end{bmatrix}}.
\end{equation}
The covariance matrix $\mathbf{\Sigma}_\mathbf{h}$ of $\mathbf{h}$ is a Toeplitz matrix, which is illustrated with the following small example more compactly:
\begin{center}
\line(1,0){450}
\end{center}
\begin{example}
In this example, we assume that $M=4$. Thus, we obtain for the vector $\mathbf{h}$ of collected channel coefficients of the observation interval $\mathcal{I}_M$:
\begin{equation}
\mathbf{h} = [h[3], h[2], h[1], h[0]]^{T},
\end{equation}
Accordingly, the covariance matrix of $\mathbf{h}$ is:
\begin{equation}
\mathbf{\Sigma}_\mathbf{h} =
\begin{bmatrix}
R_h[0]   & R_h[1]   & R_h[2] & R_h[3] \\  
R_h^*[1] & R_h[0]   & R_h[1] & R_h[2] \\
R_h^*[2] & R^*_h[1] & R_h[0] & R_h[1] \\
R_h^*[3] & R^*_h[2] & R^*_h[1] & R_h[0] \\
\end{bmatrix}.
\end{equation}
\end{example}
\begin{center}
\line(1,0){450}
\end{center}
\section{LMMSE Predictor}

With the physical propagation channel model explained above, the goal is now to predict channel coefficients of the prediction interval $\mathcal{I}_N$ given noisy observations of the channel coefficients of the observation interval $\mathcal{I}_M$. The noisy observations of channel coefficients within the observation interval $\mathcal{I}_M$ are collected in a vector $\mathbf{y}$ \cite{Zemen, Zemen3}:
\begin{equation}
    \mathbf{y} = \mathbf{h} + \mathbf{n},
    \label{noisyObs}
\end{equation}
where the complex additive white Gaussian noise (AWGN) is described by:
\begin{equation}
        \mathbf{n} \sim \mathcal{N}_\mathbb{C}(\mathbf{0},\mathbf{\Sigma}_\mathbf{n} = \sigma_n^2 \mathbf{I}_M),
\end{equation}
with $\mathbf{I}_M$ being the $M\times M$ identity matrix. The signal-to-noise ratio (SNR) is SNR $= {1/ \sigma_n^2}$.
Accordingly, the covariance matrix of the noisy observations $\mathbf{y}$ is \cite{Zemen, Zemen3}:
\begin{equation}
        \mathbf{\Sigma}_\mathbf{y} = \mathbf{\Sigma}_\mathbf{h}  + \sigma_n^2 \mathbf{I}_M.
        \label{covmatnoisy}
\end{equation}
With these quantities in hand, channel coefficients of the prediction interval $\mathcal{I}_N$ can be obtained with the $l-$step LMMSE predictor~\cite{Zemen,Kay}:
\begin{equation}
\hat{h}[m] = \hat{h}_m = \mathbf{c}_{h_{m}\mathbf{y}}^{H} \mathbf{\Sigma}_{\mathbf{y}}^{-1} \mathbf{y},
\label{lstepWP}
\end{equation}
with $m \in \mathcal{I}_N$ and $l = m-(M-1)$ and the correlation vector $\mathbf{c}_{h_{m}\mathbf{y}}^{H}$ of following form:
\begin{equation}
\mathbf{c}_{h_{m}\mathbf{y}}^{H} = [R_h[l], R_h[1+l], \dots, R_h[M-1+l]],
\label{corrvec}
\end{equation}
where the entries of the correlation vector $\mathbf{c}_{h_{m}\mathbf{y}}^{H}$ are for a finite number of propagation paths $P$ according to \eqref{coorelationFunc_finiteP} and for $P\to\infty$ according to \eqref{coorelationFunc_infP}.\newline
\\
In the following, the LMMSE predictor is reformulated in order to derive other predictors, which are described in the following sections. The reformulated version is obtained with two steps:
\begin{itemize}
    \item First, the covariance matrix $\mathbf{\Sigma}_\mathbf{h}$ of the channel observations is extended by the desired step length $l$, for which a predicted channel coefficient should be obtained. 
    \item Then, by making use of the structure of the extended covariance matrix $\mathbf{\Sigma}_\mathbf{h}^{l-\text{ext}}$, it is possible to identify the covariance matrix $\mathbf{\Sigma}_\mathbf{h}$ of the channel observations and the correlation vector $\mathbf{c}_{h_{m}\mathbf{y}}^{H}$ as specific parts of the extended covariance matrix $\mathbf{\Sigma}_\mathbf{h}^{l-\text{ext}}$.
\end{itemize}
These two basic steps are explained in the following. For a fixed step length $l$ the vector of channel coefficients $\mathbf{h}$ of the observation interval $\mathcal{I}_M$ is artificially extended by $l$ channel coefficients of the prediction interval $\mathcal{I}_N$:
\begin{eqnarray}
    \mathbf{h}^{l-\text{ext}}
      &=&   [\textcolor{red}{h[m]},\dots,\textcolor{red}{h[M]} ,\mathbf{h}^T]^T \\ 
      &=& [\textcolor{red}{h[m]},\dots,\textcolor{red}{h[M]}, h[M-1], \dots, h[1], h[0]]^{T},
\end{eqnarray}
with $m \in \mathcal{I}_N$ and $m = l+(M-1)$. Therefore, the extended covariance matrix $\mathbf{\Sigma}_\mathbf{h}^{l-\text{ext}}$ is of following form:
\begin{equation}
\small
\mathbf{\Sigma}_\mathbf{h}^{l-\text{ext}} =
\begin{bmatrix}
\textcolor{red}{R_h[0]}   & \textcolor{red}{R_h[1]}   & \textcolor{red}{R_h[2]}   & \textcolor{red}{\dots} & \textcolor{red}{R_h[m-1]} & \textcolor{red}{R_h[m]} \\  
\textcolor{red}{R^*_h[1]}   & \textcolor{red}{\ddots}  &   &  & &  \textcolor{red}{\vdots}\\ 
\textcolor{red}{R^*_h[2]}   &   & \textcolor{red}{R_h[0]}   & \textcolor{red}{R_h[1]} & \textcolor{red}{\dots} & \textcolor{red}{R_h[M]} \\ 
\textcolor{red}{\vdots} &  & \textcolor{red}{R^*_h[1]}   & R_h[0] & \dots & {R_h[M-1]} \\
\textcolor{red}{R^*_h[m-1]} &  & \textcolor{red}{\vdots}   & \vdots & \ddots & \vdots \\
\textcolor{red}{R^*_h[m]} & \textcolor{red}{\dots} & \textcolor{red}{R^*_h[M]}   & R^*_h[M-1] & \dots & {R_h[0]}
\end{bmatrix}.
\end{equation}
We can readily identify that the entries of the correlation vector $\mathbf{c}_{h_{m}\mathbf{y}}^{H}$ from \eqref{corrvec} and the covariance matrix $\mathbf{\Sigma}_\mathbf{h}$ of the vector of channel coefficients $\mathbf{h}$ of the observation interval $\mathcal{I}_M$ are embedded in the extended covariance matrix $\mathbf{\Sigma}_\mathbf{h}^{l-\text{ext}}$. The correlation vector $\mathbf{c}_{h_{m}\mathbf{y}}^{H}$ is identical to the zeroth row starting from the $l-$th column of the extended covariance matrix $\mathbf{\Sigma}_\mathbf{h}^{l-\text{ext}}$ and the covariance matrix $\mathbf{\Sigma}_\mathbf{h}$ is by construction the bottom right part of the extended covariance matrix $\mathbf{\Sigma}_\mathbf{h}^{l-\text{ext}}$. Which is illustrated again with a small example: 
\begin{center}
\line(1,0){450}
\end{center}
\begin{example}
In this example we again assume that $M=4$ and wish to predict the channel coefficient following the $M-1$-th channel coefficient. In this way, we have a single-step predictor, thus $l=1$. We obtain for the vector $\mathbf{h}$ of collected channel coefficients of the observation interval $\mathcal{I}_M$ by extending the  vector by one channel coefficient:
\begin{equation}
\mathbf{h}^{l-\text{ext}} = [\textcolor{red}{h[4]},h[3], h[2], h[1], h[0]]^{T}.
\end{equation}
Accordingly, the extended covariance matrix $\mathbf{\Sigma}_\mathbf{h}^{l-\text{ext}}$ of $\mathbf{h}^{l-\text{ext}}$ is:
\begin{equation}
\mathbf{\Sigma}_\mathbf{h}^{l-\text{ext}} =
\begin{bmatrix}
\textcolor{red}{R_h[0]}   & \textcolor{red}{R_h[1]}   & \textcolor{red}{R_h[2]} & \textcolor{red}{R_h[3]} & \textcolor{red}{R_h[4]} \\  
\textcolor{red}{R_h^*[1]} & R_h[0]   & R_h[1] & R_h[2] & R_h[3] \\
\textcolor{red}{R_h^*[2]} & R^*_h[1] & R_h[0] & R_h[1] & R_h[2] \\
\textcolor{red}{R_h^*[3]} & R^*_h[2] & R^*_h[1] & R_h[0] & R_h[1] \\
\textcolor{red}{R_h^*[4]} & R^*_h[3] & R^*_h[2] & R^*_h[1] & R_h[0] \\
\end{bmatrix},
\end{equation}
where we have extended the covariance matrix by one row and one column. We can identify that the correlation vector $\mathbf{c}_{h_{m}\mathbf{y}}^{H}$ is identical to the zeroth row starting from the first column of the extended covariance matrix $\mathbf{\Sigma}_\mathbf{h}^{l-\text{ext}}$:
\begin{equation}
\mathbf{\Sigma}_\mathbf{h}^{l-\text{ext}} =
\begin{bmatrix}
R_h[0]   & \textcolor{red}{R_h[1]}   & \textcolor{red}{R_h[2]} & \textcolor{red}{R_h[3]} & \textcolor{red}{R_h[4]} \\  
R_h^*[1] & R_h[0]   & R_h[1] & R_h[2] & R_h[3] \\
R_h^*[2] & R^*_h[1] & R_h[0] & R_h[1] & R_h[2] \\
R_h^*[3] & R^*_h[2] & R^*_h[1] & R_h[0] & R_h[1] \\
R_h^*[4] & R^*_h[3] & R^*_h[2] & R^*_h[1] & R_h[0] \\
\end{bmatrix}.
\end{equation}
The covariance matrix $\mathbf{\Sigma}_\mathbf{h}$ is by construction embedded at the bottom right part of the extended covariance matrix $\mathbf{\Sigma}_\mathbf{h}^{l-\text{ext}}$:
\begin{equation}
\mathbf{\Sigma}_\mathbf{h}^{l-\text{ext}} =
\begin{bmatrix}
R_h[0]   & R_h[1]   & R_h[2] & R_h[3] & R_h[4] \\  
R_h^*[1] & \textcolor{red}{R_h[0]}   & \textcolor{red}{R_h[1]} & \textcolor{red}{R_h[2]} & \textcolor{red}{R_h[3]} \\
R_h^*[2] & \textcolor{red}{R^*_h[1]} & \textcolor{red}{R_h[0]} & \textcolor{red}{R_h[1]} & \textcolor{red}{R_h[2]} \\
R_h^*[3] & \textcolor{red}{R^*_h[2]} & \textcolor{red}{R^*_h[1]} & \textcolor{red}{R_h[0]} & \textcolor{red}{R_h[1]} \\
R_h^*[4] & \textcolor{red}{R^*_h[3]} & \textcolor{red}{R^*_h[2]} & \textcolor{red}{R^*_h[1]} & \textcolor{red}{R_h[0]} \\
\end{bmatrix}.
\end{equation}
\end{example}
\begin{center}
\line(1,0){450}
\end{center}
With the following definitions:
\begin{equation}
       \mathbf{e}_1^T = [1,0, \dots, 0] \hspace{1.5cm} (1 \times M+l)
\end{equation}
\begin{equation}
       \mathbf{S} = 
       \begin{bmatrix}
        \mathbf{0} \\
        \mathbf{I}_M
\end{bmatrix} \hspace{1.5cm} (M+l \times M).
\end{equation}
the correlation vector $\mathbf{c}_{h_{m}\mathbf{y}}^{H}$ and the covariance matrix can be extracted $\mathbf{\Sigma}_\mathbf{h}$ from the extended covariance matrix $\mathbf{\Sigma}_\mathbf{h}^{l-\text{ext}}$ by:
\begin{equation}
       { \mathbf{c}_{h_{m}\mathbf{y}}^{H} } = \mathbf{e}_1^T \mathbf{\Sigma}_\mathbf{h}^{l-\text{ext}} \mathbf{S}
       \label{corrvecreformu}
\end{equation}
\begin{equation}
       {\mathbf{\Sigma}_\mathbf{h}} = \mathbf{S}^T \mathbf{\Sigma}_\mathbf{h}^{l-\text{ext}} \mathbf{S}.
       \label{covmatreformu}
\end{equation}
With \eqref{covmatreformu} the expression for the covariance matrix of the noisy observations $ \mathbf{\Sigma}_\mathbf{y}$ from \eqref{covmatnoisy} is rewritten to:
\begin{equation}
        \mathbf{\Sigma}_\mathbf{y} = \mathbf{S}^T \mathbf{\Sigma}_\mathbf{h}^{l-\text{ext}} \mathbf{S} + \sigma_n^2 \mathbf{I}_M.
        \label{covmatnoisyreformu}
\end{equation}
\begin{center}
\line(1,0){450}
\end{center}
\begin{example}
According to the previous example, $M=4$ and $l=1$. Thus, we obtain:
\begin{equation}
       \mathbf{e}_1^T = [1, 0, 0, 0, 0] \hspace{1.5cm} (1 \times 5)
\end{equation}
\begin{equation}
       \mathbf{S} = 
       \begin{bmatrix}
        \mathbf{0}^T \\
        \mathbf{I}_4
\end{bmatrix} \hspace{1.5cm} (5 \times 4).
\end{equation}
\end{example}
\begin{center}
\line(1,0){450}
\end{center}
With the covariance matrix extension and the extraction of the respective components in hand, we are now able to reformulate the $l-$step LMMSE predictor from \eqref{lstepWP} by incorporating our results from \eqref{corrvecreformu} and \eqref{covmatnoisyreformu}:
\begin{align}
    \hat{h}_m 
      & = \mathbf{e}_1^T \mathbf{\Sigma}_\mathbf{h}^{l-\text{ext}} \mathbf{S} (\mathbf{S}^T \mathbf{\Sigma}_\mathbf{h}^{l-\text{ext}} \mathbf{S} +  \sigma_n^2 \mathbf{I}_M)^{-1} \mathbf{y}
      \label{lstepWPreformu} \\
      & = \mathbf{e}_1^T\mathbf{W}^{l-\text{ext}}\mathbf{y}. 
      \label{lstepWPreformu2}
\end{align}

\section{Gridded Predictor}

In the following, we use Bayes' approach of \cite{Neumann} to derive an approximated MMSE predictor. The proposed solution is based on the random variables ${\boldsymbol{\delta}}$ described by a given distribution $p({\boldsymbol{\delta}})$ and the assumption that for each sample of ${\boldsymbol{\delta}}$ the closed-form solution $ {\mathbf{W}}_{\boldsymbol{\delta}} $ of the LMMSE predictor according to $ \mathbf{W}^{l-\text{ext}} $ as in \eqref{lstepWPreformu2} is available \cite{Neumann, Koller, Hellings}:
\begin{equation}
       \hat{\mathbf{W}}_\text{MMSE} = \int p({\boldsymbol{\delta}}|\mathbf{y}) {\mathbf{W}}_{\boldsymbol{\delta}} d{\boldsymbol{\delta}}.
\end{equation}
Note that each realization of the random ${\boldsymbol{\delta}}$ corresponds to the DoAs of a sampled scenario that determines the path-specific Doppler shift.
By using Bayes' theorem:
\begin{equation}
        p({\boldsymbol{\delta}}|\mathbf{y}) = \dfrac{p(\mathbf{y}|{\boldsymbol{\delta}}) p({\boldsymbol{\delta}})} { \int p(\mathbf{y}|{\boldsymbol{\delta}}) p( {\boldsymbol{\delta}}) d{\boldsymbol{\delta}} },
\end{equation}
it follows for the estimated filter:
\begin{equation}
       \hat{\mathbf{W}}_\text{MMSE} = \dfrac{\int p(\mathbf{y}|{\boldsymbol{\delta}}) {\mathbf{W}}_{\boldsymbol{\delta}} p({\boldsymbol{\delta}})  d{\boldsymbol{\delta}} } { \int p(\mathbf{y}|{\boldsymbol{\delta}}) p( {\boldsymbol{\delta}}) d{\boldsymbol{\delta}} }.
       \label{estFilt}
\end{equation}
The likelihood of $\mathbf{y}$ given ${\boldsymbol{\delta}}$ is assumed to be Gaussian:\footnote{We now indexed the second order statistical moments with $\boldsymbol{\delta}$ to express the dependency on the selected sample of the scenario.}
\begin{equation}
  p(\mathbf{y}|{\boldsymbol{\delta}}) \propto \dfrac{\exp{(-\mathbf{y}^H\mathbf{\Sigma}_{\mathbf{y}_{\boldsymbol{\delta}}}^{-1}\mathbf{y})}}{|\mathbf{\Sigma}_{\mathbf{y}_{\boldsymbol{\delta}}}|},
\end{equation}
which we can rewrite to:
\begin{equation}
  p(\mathbf{y}|{\boldsymbol{\delta}}) \propto \exp{(-\mathrm{tr}(\mathbf{\Sigma}_{\mathbf{y}_{\boldsymbol{\delta}}}^{-1} \mathbf{y}\mathbf{y}^H))}{|\mathbf{\Sigma}_{\mathbf{y}_{\boldsymbol{\delta}}}^{-1}|},
\end{equation}
by making use of some basic linear algebra rules. We now wish to express $\mathbf{\Sigma}_{\mathbf{y}_{\boldsymbol{\delta}}}^{-1}$ in terms of ${\mathbf{W}}_{\boldsymbol{\delta}}$. To this end, we firstly identify ${\mathbf{W}}_{\boldsymbol{\delta}}$ with the predictor in \eqref{lstepWPreformu}:
\begin{equation}
       \mathbf{\Sigma}^{l-\text{ext}}_{\mathbf{h}_{\boldsymbol{\delta}}} \mathbf{S} (\mathbf{S}^T \mathbf{\Sigma}^{l-\text{ext}}_{\mathbf{h}_{\boldsymbol{\delta}}} \mathbf{S} + \mathbf{\Sigma}_\mathbf{n})^{-1} = {\mathbf{W}}_{\boldsymbol{\delta}} 
\end{equation}
\begin{equation}
       \mathbf{\Sigma}^{l-\text{ext}}_{\mathbf{h}_{\boldsymbol{\delta}}} \mathbf{S}  = {\mathbf{W}}_{\boldsymbol{\delta}} (\mathbf{S}^T \mathbf{\Sigma}^{l-\text{ext}}_{\mathbf{h}_{\boldsymbol{\delta}}} \mathbf{S} + \mathbf{\Sigma}_\mathbf{n})
\end{equation}
\begin{equation}
      \mathbf{S}^T \mathbf{\Sigma}^{l-\text{ext}}_{\mathbf{h}_{\boldsymbol{\delta}}} \mathbf{S}  = \mathbf{S}^T {\mathbf{W}}_{\boldsymbol{\delta}} (\mathbf{S}^T \mathbf{\Sigma}^{l-\text{ext}}_{\mathbf{h}_{\boldsymbol{\delta}}} \mathbf{S} + \mathbf{\Sigma}_\mathbf{n})
\end{equation}
\begin{equation}
      \mathbf{S}^T \mathbf{\Sigma}^{l-\text{ext}}_{\mathbf{h}_{\boldsymbol{\delta}}} \mathbf{S} + \mathbf{\Sigma}_\mathbf{n}  = \mathbf{S}^T {\mathbf{W}}_{\boldsymbol{\delta}} (\mathbf{S}^T \mathbf{\Sigma}^{l-\text{ext}}_{\mathbf{h}_{\boldsymbol{\delta}}} \mathbf{S} + \mathbf{\Sigma}_\mathbf{n}) + \mathbf{\Sigma}_\mathbf{n}
\end{equation}
\begin{equation}
     \mathbf{I}_M  = \mathbf{S}^T {\mathbf{W}}_{\boldsymbol{\delta}} + \mathbf{\Sigma}_\mathbf{n} (\mathbf{S}^T \mathbf{\Sigma}_\mathbf{h}^{l-\text{ext}} \mathbf{S} + \mathbf{\Sigma}_\mathbf{n})^{-1}
\end{equation}
\begin{equation}
     \mathbf{\Sigma}_{\mathbf{y}_{\boldsymbol{\delta}}}^{-1} = \mathbf{\Sigma}_\mathbf{n}^{-1}  (\mathbf{I}_M - \mathbf{S}^T {\mathbf{W}}_{\boldsymbol{\delta}}).
\end{equation}
The likelihood is now re-expressed in terms of ${\mathbf{W}}_{\boldsymbol{\delta}}$ by:
\begin{equation}
  p(\mathbf{y}|{\boldsymbol{\delta}}) \propto \exp{(-\mathrm{tr}(\mathbf{\Sigma}_\mathbf{n}^{-1}  (\mathbf{I}_M - \mathbf{S}^T {\mathbf{W}}_{\boldsymbol{\delta}}) \mathbf{y}\mathbf{y}^H))}{|\mathbf{\Sigma}_\mathbf{n}^{-1}  (\mathbf{I}_M - \mathbf{S}^T {\mathbf{W}}_{\boldsymbol{\delta}})|}.
\end{equation}
Since the noise covariance matrix $\mathbf{\Sigma}_\mathbf{n}$ does not depend on ${\boldsymbol{\delta}}$, we can further simplify the likelihood $p(\mathbf{y}|{\boldsymbol{\delta}})$:
\begin{equation}
  p(\mathbf{y}|{\boldsymbol{\delta}}) \propto \exp{(\mathrm{tr}(\mathbf{\Sigma}_\mathbf{n}^{-1} \mathbf{S}^T {\mathbf{W}}_{\boldsymbol{\delta}} \mathbf{y}\mathbf{y}^H))}|{\mathbf{I}_M - \mathbf{S}^T {\mathbf{W}}_{\boldsymbol{\delta}}|}.
\end{equation}
By defining:
\begin{eqnarray}
    \hat{\mathbf{C}} 
      &=& \dfrac{1}{\sigma^2_n}\mathbf{y}\mathbf{y}^H \\ 
      b_{\boldsymbol{\delta}}
      &=&  \mathrm{log}|{\mathbf{I}_M - \mathbf{S}^T {\mathbf{W}}_{\boldsymbol{\delta}}|},
      \label{biasTerm}
\end{eqnarray}
the likelihood $p(\mathbf{y}|{\boldsymbol{\delta}})$ is reformulated to:
\begin{equation}
  p(\mathbf{y}|{\boldsymbol{\delta}}) \propto \exp{(\mathrm{tr}( \mathbf{S}^T {\mathbf{W}}_{\boldsymbol{\delta}} \hat{\mathbf{C}})}+b_{\boldsymbol{\delta}}).
\end{equation}
We can now incorporate the result for $p(\mathbf{y}|{\boldsymbol{\delta}})$ into \eqref{estFilt}:
\begin{equation}
\hat{\mathbf{W}}_\text{MMSE} =  \dfrac{\int \exp{(\mathrm{tr}( \mathbf{S}^T {\mathbf{W}}_{\boldsymbol{\delta}} \hat{\mathbf{C}})}+b_{\boldsymbol{\delta}}) {\mathbf{W}}_{\boldsymbol{\delta}} p({\boldsymbol{\delta}})  d{\boldsymbol{\delta}} } { \int \exp{(\mathrm{tr}( \mathbf{S}^T {\mathbf{W}}_{\boldsymbol{\delta}} \hat{\mathbf{C}})}+b_{\boldsymbol{\delta}}) p( {\boldsymbol{\delta}}) d{\boldsymbol{\delta}} }.
\end{equation}
Analogous to \eqref{lstepWPreformu}, the approximated MMSE predictor is: 
\begin{equation}
    \hat{\mathbf{w}}^T(\hat{\mathbf{C}}) 
      =  \mathbf{e}_1^T \dfrac{\int \exp{(\mathrm{tr}( \mathbf{S}^T {\mathbf{W}}_{\boldsymbol{\delta}} \hat{\mathbf{C}})}+b_{\boldsymbol{\delta}}) {\mathbf{W}}_{\boldsymbol{\delta}} p({\boldsymbol{\delta}})  d{\boldsymbol{\delta}} } { \int \exp{(\mathrm{tr}( \mathbf{S}^T {\mathbf{W}}_{\boldsymbol{\delta}} \hat{\mathbf{C}})}+b_{\boldsymbol{\delta}}) p( {\boldsymbol{\delta}}) d{\boldsymbol{\delta}} }.
      \label{filterVector}
\end{equation}

For arbitrary prior distributions $p({\boldsymbol{\delta}})$ this filter cannot be evaluated in closed form \cite{Neumann}. Similar as in \cite{Neumann}, to have a computable expression we make the following assumption:\newline
\begin{assumption}
The prior $p({\boldsymbol{\delta}})$ is discrete and uniform:
\begin{equation}
    p({\boldsymbol{\delta}_i}) = \dfrac{1}{N},  \forall i = 1,\dots N.
\end{equation}
\end{assumption}
By making this assumption the prior in \eqref{filterVector} can be replaced by $\dfrac{1}{N}$ and the integrals are replaced by sums. We end up with the Gridded Predictor:
\begin{equation}
\hat{\mathbf{w}}^T(\hat{\mathbf{C}})  = \mathbf{e}_1^T \dfrac{ \dfrac{1}{N}\sum_{i=1}^{N} \exp{(\mathrm{tr}( \mathbf{S}^T {\mathbf{W}}_{\boldsymbol{\delta}_i} \hat{\mathbf{C}})}+b_{\boldsymbol{\delta}_i}) {\mathbf{W}}_{\boldsymbol{\delta}_i}} { \dfrac{1}{N} \sum_{i=1}^{N} \exp{(\mathrm{tr}( \mathbf{S}^T {\mathbf{W}}_{\boldsymbol{\delta}_i} \hat{\mathbf{C}})}+b_{\boldsymbol{\delta}_i})},
\label{ass1Filter}
\end{equation}
where the filter of each sample ${\mathbf{W}}_{\boldsymbol{\delta}_i}$ is calculated by \eqref{lstepWPreformu} and $b_{\boldsymbol{\delta}_i}$ is evaluated by \eqref{biasTerm}. With larger $N$ the approximation error decreases. Nevertheless, there will be a gap compared to the LMMSE predictor with perfect knowledge of the statistical moments of second order based on the coveriance function $ R_h[k]$, because of a finite $N$. This gap is even more significant if the $\boldsymbol{\delta}_i$ are sampled from a prior $p({\boldsymbol{\delta}})$, with more than one propagation path. In such a case, many combinations of DoAs are possible, which can not be fully captured by the Gridded Predictor with a finite number of samples $N$.

\section{Structured Predictor}

With the Gridded Predictor it is possible to achieve prediction without the knowledge of the true PSD of the channel coefficients. The drawbacks of this predictor are its numerical complexity and a large memory requirement, due to the storage of a filter for each sample ${\mathbf{W}}_{\boldsymbol{\delta}_i}$. By making another assumption, it is possible to simplify the predictor and reduce the memory overhead:
\begin{assumption}
$\forall i = 1,\dots N$ the filters $\mathbf{S}^T {\mathbf{W}}_{\boldsymbol{\delta}_i}$ can be decomposed as:
\begin{equation}
\mathbf{S}^T {\mathbf{W}}_{\boldsymbol{\delta}_i} = \mathbf{Q}^H \mathrm{diag}(\mathbf{w}_{\boldsymbol{\delta}_i}) \mathbf{Q},
\label{ass2}
\end{equation}
with $ \mathbf{w}_{\boldsymbol{\delta}_i} \in \mathbb{R}^K$ and a common matrix $\mathbf{Q} \in \mathbb{C}^{K \times M}$.
\end{assumption}
Instead of storing a matrix for each sample ${\mathbf{W}}_{\boldsymbol{\delta}_i}$, it is now sufficient to store a vector $\mathbf{w}_{\boldsymbol{\delta}_i}$ for each sample, which reduces the memory overhead. Similar as in \cite{Neumann}, possible candidates for $\mathbf{Q}$ are \cite{Gray2006}:
\begin{itemize}
    \item The DFT matrix: $\mathbf{Q} = \mathbf{F}_1 \in  \mathbb{C}^{M \times M}$
    \newline $\,\to\,$ Circulant approximation
    \item First M columns of the $2M\times 2M$ DFT matrix: $\mathbf{Q} = \mathbf{F}_2 \in  \mathbb{C}^{2M \times M}$
    \newline $\,\to\,$ Toeplitz approximation
\end{itemize}
By defining 
\begin{equation}
\hat{\mathbf{c}} = \dfrac{1}{\sigma^2_n} |\mathbf{Qy}|^2   
\label{smallchat}
\end{equation}
and using \eqref{ass2}, it follows for the trace expressions in \eqref{ass1Filter}:
\begin{eqnarray}
    \mathrm{tr}( \mathbf{S}^T {\mathbf{W}}_{\boldsymbol{\delta}_i} \hat{\mathbf{C}})
      &=&  \mathrm{tr}( \mathbf{Q}^H \mathrm{diag}(\mathbf{w}_{\boldsymbol{\delta}_i}) \mathbf{Q} \dfrac{1}{\sigma^2_n}\mathbf{y}\mathbf{y}^H) \\ 
      &=& \mathrm{tr}( \mathrm{diag}(\mathbf{w}_{\boldsymbol{\delta}_i}) \dfrac{1}{\sigma^2_n} \mathbf{Q} \mathbf{y}\mathbf{y}^H \mathbf{Q}^H ) \\
      &=& \mathbf{w}^T_{\boldsymbol{\delta}_i} \hat{\mathbf{c}},
\end{eqnarray}
since $\hat{\mathbf{c}}$ contains the diagonal entries of the matrix
\begin{equation}
    \dfrac{1}{\sigma^2_n} \mathbf{Q} \mathbf{y}\mathbf{y}^H \mathbf{Q}^H.
\end{equation}
After reformulating the trace expressions, the Gridded Predictor from \eqref{ass1Filter} simplifies to:
\begin{equation}
\hat{\mathbf{w}}^T(\hat{\mathbf{c}})  = \dfrac{ \dfrac{1}{N}\sum_{i=1}^{N} \exp{(\mathbf{w}^T_{\boldsymbol{\delta}_i} \hat{\mathbf{c}}}+b_{\boldsymbol{\delta}_i}) \mathbf{e}_1^T {\mathbf{W}}_{\boldsymbol{\delta}_i}} { \dfrac{1}{N} \sum_{i=1}^{N} \exp{( \mathbf{w}^T_{\boldsymbol{\delta}_i} \hat{\mathbf{c}}}+b_{\boldsymbol{\delta}_i})}.
\end{equation}
We end up with the \textit{Structured Predictor} of following form:
\begin{equation}
\hat{\mathbf{w}}(\hat{\mathbf{c}})  = \mathbf{A}_2 \dfrac{ \exp{(\mathbf{A}_1\hat{\mathbf{c}}+ \mathbf{b} )} } { \mathbf{1}^T\exp{(\mathbf{A}_1\hat{\mathbf{c}}+ \mathbf{b})}},
\label{StrucPred}
\end{equation}
where the matrices $\mathbf{A}_1$ and $\mathbf{A}_2$ and the vector $\mathbf{b}$ are defined as follows:
\begin{equation}
    \mathbf{A}_1 = 
       \begin{bmatrix}
        \mathbf{w}^T_{\boldsymbol{\delta}_1} \\
        \mathbf{w}^T_{\boldsymbol{\delta}_2} \\
        \vdots \\
        \mathbf{w}^T_{\boldsymbol{\delta}_N}
\end{bmatrix} \hspace{0.5cm}
\mathbf{A}_2 = 
\begin{bmatrix}
        \mathbf{e}_1^T\mathbf{W}_{\boldsymbol{\delta}_1} \\
        \mathbf{e}_1^T\mathbf{W}_{\boldsymbol{\delta}_2} \\
        \vdots \\
        \mathbf{e}_1^T\mathbf{W}_{\boldsymbol{\delta}_N}
\end{bmatrix}^T \hspace{0.5cm}
\mathbf{b} = 
\begin{bmatrix}
        b_{\boldsymbol{\delta}_1}\\
        b_{\boldsymbol{\delta}_2}\\
        \vdots \\
        b_{\boldsymbol{\delta}_N}
\end{bmatrix}.
\label{A1A2b}
\end{equation}

\section{Neural Network Predictor}

An expert observation shows that a feed-forward neural network with one hidden layer and the softmax activation function,
\begin{equation}
 \Phi(\mathbf{x}) = \dfrac{ \exp{( \mathbf{x} )} } { \mathbf{1}^T\exp{(\mathbf{x})}}
\end{equation} has the same structure as the Structured Predictor. We define the feed-forward neural network as:
\begin{equation}
\hat{\mathbf{w}}_{NN}(\hat{\mathbf{c}})  = \mathbf{A}_{(2)} \dfrac{ \exp{(\mathbf{A}_{(1)}\hat{\mathbf{c}}+ \mathbf{b}_{(1)} )} } { \mathbf{1}^T\exp{(\mathbf{A}_{(1)}\hat{\mathbf{c}}+ \mathbf{b}_{(1)})}} + \mathbf{b}_{(2)}, 
\end{equation}
which can be visualized as the block diagram shown in the following figure:
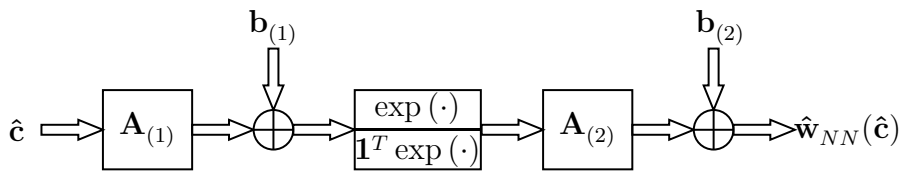
\begin{figure}[h]
\centering
    \begin{center}
    \end{center}

    \tikzset{every picture/.style={line width=0.75pt}} 
    
    \begin{tikzpicture}[x=0.75pt,y=0.75pt,yscale=-1,xscale=1]
    
    \draw   (125,105.86) -- (170,105.86) -- (170,145.86) -- (125,145.86) -- cycle ;
    \draw   (94,123.37) -- (112.6,123.37) -- (112.6,120.86) -- (125,125.89) -- (112.6,130.93) -- (112.6,128.41) -- (94,128.41) -- cycle ;
    \draw   (201,125.89) .. controls (201,120.38) and (205.48,115.91) .. (211,115.91) .. controls (216.52,115.91) and (221,120.38) .. (221,125.89) .. controls (221,131.41) and (216.52,135.87) .. (211,135.87) .. controls (205.48,135.87) and (201,131.41) .. (201,125.89) -- cycle ; \draw   (201,125.89) -- (221,125.89) ; \draw   (211,115.91) -- (211,135.87) ;
    \draw   (170,123.37) -- (188.6,123.37) -- (188.6,120.86) -- (201,125.89) -- (188.6,130.93) -- (188.6,128.41) -- (170,128.41) -- cycle ;
    \draw   (221,123.37) -- (239.6,123.37) -- (239.6,120.86) -- (252,125.89) -- (239.6,130.93) -- (239.6,128.41) -- (221,128.41) -- cycle ;
    \draw   (252,105.86) -- (315.6,105.86) -- (315.6,145.86) -- (252,145.86) -- cycle ;
    \draw   (316,123.37) -- (334.6,123.37) -- (334.6,120.86) -- (347,125.89) -- (334.6,130.93) -- (334.6,128.41) -- (316,128.41) -- cycle ;
    \draw   (347,105.86) -- (392,105.86) -- (392,145.86) -- (347,145.86) -- cycle ;
    \draw   (392,123.37) -- (410.6,123.37) -- (410.6,120.86) -- (423,125.89) -- (410.6,130.93) -- (410.6,128.41) -- (392,128.41) -- cycle ;
    \draw   (423,125.89) .. controls (423,120.38) and (427.48,115.91) .. (433,115.91) .. controls (438.52,115.91) and (443,120.38) .. (443,125.89) .. controls (443,131.41) and (438.52,135.87) .. (433,135.87) .. controls (427.48,135.87) and (423,131.41) .. (423,125.89) -- cycle ; \draw   (423,125.89) -- (443,125.89) ; \draw   (433,115.91) -- (433,135.87) ;
    \draw   (443,123.37) -- (461.6,123.37) -- (461.6,120.86) -- (474,125.89) -- (461.6,130.93) -- (461.6,128.41) -- (443,128.41) -- cycle ;
    \draw   (435.52,84.91) -- (435.52,103.51) -- (438.04,103.51) -- (433,115.91) -- (427.96,103.51) -- (430.48,103.51) -- (430.48,84.91) -- cycle ;
    \draw   (213.52,84.91) -- (213.52,103.51) -- (216.04,103.51) -- (211,115.91) -- (205.96,103.51) -- (208.48,103.51) -- (208.48,84.91) -- cycle ;
    
    \draw (147.5,125.86) node  [align=left] {$\displaystyle \mathbf{A}^{
    	}_{(1)}$};
    \draw (369.5,125.86) node  [align=left] {$\displaystyle \mathbf{A}^{
    	}_{(2)}$};
    \draw (283.5,125.89) node  [align=left] {$\displaystyle \dfrac{\exp{(\cdot)}}{\mathbf{1}^T\exp{(\cdot)}}$};
    \draw (211,74.36) node  [align=left] {$\displaystyle \mathbf{b}^{}_{(1)}$};
    \draw (436.5,74.36) node  [align=left] {$\displaystyle \mathbf{b}^{}_{(2)}$};
    \draw (81.5,125.89) node  [align=left] {$\displaystyle \mathbf{\hat{c}}$};
    \draw (500.5,125.89) node  [align=left] {$\displaystyle \mathbf{\hat{w}}^{
    	}_{NN}(\mathbf{\hat{c}})$};

    \end{tikzpicture}
    
    \caption{Feed-forward neural network with one hidden layer and softmax activation function.}
    \label{Fig2}
\end{figure}\newline
The matrix $\mathbf{A}_1$ of the Structured Predictor from \eqref{StrucPred}, which comprises the sample specific filter vectors $ \mathbf{w}_{\boldsymbol{\delta}_i} \in \mathbb{R}^K$, is equal to the weight matrix $\mathbf{A}_{(1)}$ of the first layer of the neural network:
\begin{equation}
    \mathbf{A}_{(1)} = \mathbf{A}_1.
    \label{NN1}
\end{equation}
The vector $\mathbf{b}$ is the bias vector of the first layer, thus
\begin{equation}
    \mathbf{b}_{(1)} = \mathbf{b}.
    \label{NN2}
\end{equation}
 If we carefully consider \eqref{A1A2b}, we can see that the entries of the second matrix $\mathbf{A}_2$ consist of sample specific filter vectors $\mathbf{e}_1^T{\mathbf{W}}_{\boldsymbol{\delta}_i} \in \mathbb{C}^{1\times M}$. Thus, the matrix $\mathbf{A}_2$ is complex. Therefore, we split the matrix $\mathbf{A}_2$ into its real and imaginary part and define 
\begin{equation}
     \mathbf{A}_{(2)} = 
     \begin{bmatrix}
        \Re(\mathbf{A}_2) \\
        \Im(\mathbf{A}_2) \\
\end{bmatrix}.
\label{NN3}
\end{equation}
We further define for each neuron (representing the real and imaginary parts) in the second layer a bias term, which are comprised in the vector $\mathbf{b}_{(2)}$. The Structured Predictor suggests that the bias vector of the second layer is the all zero vector:
\begin{equation}
    \mathbf{b}_{(2)} = \mathbf{0}.
    \label{NN4}
\end{equation}
Accordingly,  the output of the neural network $\hat{\mathbf{w}}_{NN}(\hat{\mathbf{c}})$ is the concatenation of the real and imaginary parts of the Structured Predictor $\hat{\mathbf{w}}(\hat{\mathbf{c}})$:
\begin{equation}
    \hat{\mathbf{w}}_{NN}(\hat{\mathbf{c}}) = 
         \begin{bmatrix}
        \hat{\mathbf{w}}_{NN,\Re}(\hat{\mathbf{c}}) \\
        \hat{\mathbf{w}}_{NN,\Im}(\hat{\mathbf{c}}) \\
\end{bmatrix} = 
\begin{bmatrix}
        \Re(\hat{\mathbf{w}}(\hat{\mathbf{c}})) \\
        \Im(\hat{\mathbf{w}}(\hat{\mathbf{c}})) \\
\end{bmatrix}.
\end{equation}

\subsection{Learning the MMSE Channel Predictor}
The idea is now to initialize the weights $\mathbf{A}_{(1)}$ and $\mathbf{A}_{(2)}$ and biases $\mathbf{b}_{(1)}$ and $\mathbf{b}_{(2)}$ of the Neural Network Predictor with the parameters of the Structured Predictor according to \eqref{NN1}-\eqref{NN4} and train the Neural Network Predictor to achieve a better performance as compared to the predictors described above. Therefore, a predefined number of mini-batches are generated. A mini-batch consists of $B$ channel realizations $\mathbf{h}_{b,\mathcal{I}_M}$, where each comprises $M$ channel coefficients of the observation interval $\mathcal{I}_M$ (see \eqref{hObsInt}), and corresponding channel coefficients $h_{b,\mathcal{I}_N}$ of the observation interval $\mathcal{I}_N$ (for the desired step $l$), with $b = 1,2,\dots,B$. For each channel realization $\mathbf{h}_{b,\mathcal{I}_M}$ a noisy version $\mathbf{y}_{b,\mathcal{I}_M}$ is generated by adding complex AWGN with known variance $\sigma_n^2$ (see \eqref{noisyObs}). According to \eqref{smallchat}, the input of the neural network $\mathbf{\hat{c}}_b$ for each $\mathbf{y}_{b,\mathcal{I}_M}$ can be evaluated depending on $\mathbf{Q}$ (Toeplitz or Circular). For each input $\mathbf{\hat{c}}_b$ a specific filter $\hat{\mathbf{w}}_{NN}(\hat{\mathbf{c}}_b)$ is present at the output, which can be further processed to obtain an estimate $\hat{h}_{b,\mathcal{I}_N}$ by calculating:
\begin{equation}
   \hat{h}_{b,\mathcal{I}_N} = [\hat{\mathbf{w}}_{NN,\Re}(\hat{\mathbf{c}}_b) + j\hat{\mathbf{w}}_{NN,\Im}(\hat{\mathbf{c}}_b)]^T\mathbf{y}_{b,\mathcal{I}_M}
\end{equation}
As performance metric (cost function), we choose the mean squared error (MSE). The stochastic gradient is then:
\begin{equation}
    \mathbf{g} = \dfrac{1}{B} \sum_{b=1}^{B} \dfrac{\partial}{\partial[\mathbf{A}_{(i)};\mathbf{b}_{(i)}] } \| h_{b,\mathcal{I}_N} - \hat{h}_{b,\mathcal{I}_N}\|_2^2,
\end{equation}
with $i = 1,2$. Then, the variables of the neural network are updated with a desired gradient algorithm (e.g., \cite{Kingma}). The described procedure is repeated until a convergence criterion is fulfilled. This learning procedure is described more compactly in the following:
\begin{algorithm}[H]
\caption{Learning the MMSE Channel Predictor}
\begin{algorithmic}[1]
\STATE Initialize the Neural Network with the Structured Predictor as described in \eqref{NN1}-\eqref{NN4}.
\STATE Generate a mini-batch of in total $B$ channel realizations, of the observation interval $\mathbf{h}_{b,\mathcal{I}_M}$ and corresponding channel coefficients of the prediction interval (of desired prediction step $l$) $h_{b,\mathcal{I}_N}$, for $b = 1,2,\dots,B$.
\STATE Generate noisy version $\mathbf{y}_{b,\mathcal{I}_M}$ of $\mathbf{h}_{b,\mathcal{I}_M}$ and calculate $\mathbf{\hat{c}}_b$ (input of the neural network), for $b = 1,2,\dots,B$.
\STATE Calculate stochastic gradient:
\begin{equation*}
    \mathbf{g} = \dfrac{1}{B} \sum_{b=1}^{B} \dfrac{\partial}{\partial[\mathbf{A}_{(i)};\mathbf{b}_{(i)}] } \| h_{b,\mathcal{I}_N} - [\hat{\mathbf{w}}_{NN,\Re}(\hat{\mathbf{c}_b}) + j\hat{\mathbf{w}}_{NN,\Im}(\hat{\mathbf{c}_b})]^T\mathbf{y}_{b,\mathcal{I}_M}\|_2^2,
\end{equation*}
with $i = 1,2$.
\STATE Update the variables of the neural network with a desired gradient algorithm (e.g., \cite{Kingma})
\STATE Repeat steps 2-5 until a convergence criterion is fulfilled.
\end{algorithmic}
\end{algorithm}

\section{Simulation Results}

In this Section, we discuss the performances of the previously described predictors. As baseline we use the LMMSE predictor with perfect knowledge of the statistical moments of second order based on the coveriance function $ R_h[k]$ (\textit{LMMSE Perfect}). The LMMSE predictor with the assumption of $P \to \infty$, is denoted as \textit{LMMSE Jakes}, Clearly, assuming infinitely many paths is not true for specific cases with a finite number of paths. Nevertheless, constructing the LMMSE predictor with the assumption of having infinitely many paths is straightforward, since in this case the covariance function is equal to the zeroth order Bessel function (see \eqref{coorelationFunc_infP}). The Gridded Predictor is simply denoted as \textit{Gridded}, and the Structured Predictor as \textit{Structured Toep} (Toepltiz approx.) or as \textit{Structured Circ} (circulant approx.). The Neural Network Predictor is denoted as \textit{NN Toep} or as \textit{NN Circ}. Table \ref{tab:conpred} summarizes all considered predictors.

For all simulations in the following the symbol duration $T_s = \SI{20.57}{\micro\second}$ and the carrier frequency $f_c = \SI{2}{\giga\hertz}$ as in \cite{Zemen}. For the construction of the Gridded Predictor, the Structured Predictors and the Neural Network Predictors, a fixed number of samples is needed, which is predefined depending on the number of observed symbols $M$, in order to achieve easier interpretation in terms of computational complexity. For the cases, where $\mathbf{Q} = \mathbf{F}_2$ (Toeplitz assumption) the number of samples $N$ is doubled (the input of the Structured Predictor and the Neural Network Predictor with $\mathbf{Q} = \mathbf{F}_2$ is twice as long as for the case $\mathbf{Q} = \mathbf{F}_1$). 

All of the predictors are specifically constructed for each simulated velocity, i.e., we have to construct and train the Neural Network Predictors for each velocity separately. The performances of the predictors are evaluated by calculating the MSE of 200.000 predictions.
\begin{table}[H]
    \begin{center}
    \caption{Considered Predictors in the simulations}
    \label{tab:conpred}
    \begin{tabular}{|l|l|}
    \hline
    LMMSE Perfect & with perfect knowledge of the statistical moments of second order based on $ R_h[k]$ \\ \hline
    LMMSE Jakes & LMMSE predictor with assumption $P \to \infty$ \\ \hline
    Gridded  &  Gridded Predictor \\ \hline
    Structured Toep  & Structured Predictor with $\mathbf{Q} = \mathbf{F}_2$   \\ \hline
    Structured Circ  & Structured Predictor with $\mathbf{Q} = \mathbf{F}_1$   \\ \hline
    NN Toep  & Neural Network Predictor with $\mathbf{Q} = \mathbf{F}_2$ \\ \hline
    NN Circ  & Neural Network Predictor with $\mathbf{Q} = \mathbf{F}_1$ \\ \hline
    \end{tabular}
    \end{center}
\end{table} 
In the following simulation (Fig. \ref{fig:M16_N4_1Path_Snr10}), the number of observed symbols $M = 16$ and the prediction step $l=4$. The SNR is $10\si{dB}$ and the number of impinging plane-waves at the user is one, i.e., $P=1$. The number of samples for the construction of the predictors is set to $N=16$ or $N=32$ (depending on $\mathbf{Q}$).
\begin{figure}[H]
    \centering
    \resizebox{340pt}{200pt}{
    \begin{tikzpicture}
\begin{axis}[
legend pos= outer north east,
legend style={font=\footnotesize},
xmin=0,
xmax=100,
ymin=1e-3,
ymax=5e-2,
xlabel={\textit{v} [\si{km/h}]},
ylabel={MSE},
grid=both,
ymode=log
]
\addplot[line width=1,TUMBeamerRed,solid,mark=x]
table[x=velocity,y=mse] {M16_N4_1Path_Snr10_WienerGenie.txt};
\addlegendentry{LMMSE Perfect};
\addplot[line width=1,solid,mark=diamond]
table[x=velocity,y=mse] {M16_N4_1Path_Snr10_WienerJakes.txt};
\addlegendentry{LMMSE Jakes};
\addplot[line width=1,TUMBeamerOrange,solid,mark=Mercedes star]
table[x=velocity,y=mse] {M16_N4_1Path_Snr10_Gridded.txt};
\addlegendentry{Gridded};
\addplot[line width=1,TUMBeamerGreen,solid,mark=square]
table[x=velocity,y=mse] {M16_N4_1Path_Snr10_StructuredToep.txt};
\addlegendentry{Structured Toep};
\addplot[line width=1,TUMBeamerBlue,solid,mark=triangle]
table[x=velocity,y=mse] {M16_N4_1Path_Snr10_StructuredCirc.txt};
\addlegendentry{Structured Circ};
\addplot[line width=1,TUMBeamerLightBlue,solid,mark=star]
table[x=velocity,y=mse] {M16_N4_1Path_Snr10_NNToep.txt};
\addlegendentry{NN Toep};
\addplot[line width=1,TUMBeamerDarkRed,solid,mark=o]
table[x=velocity,y=mse] {M16_N4_1Path_Snr10_NNCirc.txt};
\addlegendentry{NN Circ};
\end{axis}
\end{tikzpicture}
}
\caption{MSE at prediction step $l=4$, $M=16$, SNR $=10\si{dB}$, $P=1$, $f_c=2\si{GHz}$, $T_{s} = 20.57\si{\micro\second}$, $N=16$ or $N=32$ }
\label{fig:M16_N4_1Path_Snr10}
\end{figure}
The MSE of the LMMSE Perfect predictor remains constant for all velocities. The LMMSE Perfect predictor outperforms the LMMSE Jakes predictor for all velocities, since the LMMSE Perfect predictor has perfect knowledge of the spectrum, whereas the LMMSE Jakes predictor assumes $P\to\infty$. As compared to the LMMSE Jakes predictor, the Structured Circ predictor performs worse, whereas the Gridded and Structured Toep predictor outperform the LMMSE Jakes predictor for velocities higher than $50\si{km/h}$. As explained above the NN Toep is initialized with the Structured Toep predictor and the NN Circ predictor is initialized with the Structured Circ predictor. For training the Neural Network Predictors 3000 mini-batches, each of size $B=50$, were used. After training the Neural Network Predictors, both of them outperform the LMMSE Jakes predictor for all considered velocities. The NN Circ predictor has a lower MSE as compared to the Gridded predictor for velocities smaller than $70\si{km/h}$. The NN Toep predictor outperforms the Gridded predictor for all velocities. We can conclude that the the Neural Network Predictors are able to compensate the approximation error of Assumption 1 with a finite number of samples $N$. Already the differences in the MSEs for the Structured Toep and Structured Circ predictors suggest that the Toeplitz assumption ($\mathbf{Q} = \mathbf{F}_2$) is a better approximation. In addition to that, the neural network size with Toeplitz assumption is twice as large as compared to the Circular case. This may also explain the gap between the two Neural Network Predictor performances.
For the velocity range from $0\si{km/h}$ to $30\si{km/h}$ (Fig. \ref{fig:M16_N4_1Path_Snr10}) the Neural Network Predictors even outperform the LMMSE Perfect predictor based on the knowledge of the covariance function. This is the consequence of the assumed channel model, i.e., channel coefficients, constructed with a low number of paths (propagation channel models with specular geometry), which are not Gaussian distributed render the LMMSE predictor to be not optimal and therefore to be outperformed by other approaches that take into account the actual underlying distribution of channel coefficients or their respective samples.
\\

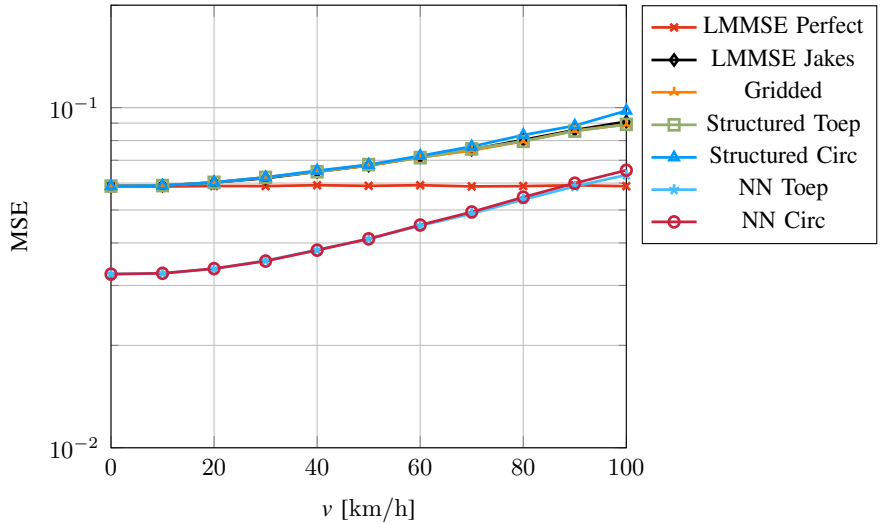
\begin{figure}[H]
    \centering
    \resizebox{340pt}{200pt}{
    \begin{tikzpicture}
\begin{axis}[
legend pos= outer north east,
legend style={font=\footnotesize},
xmin=0,
xmax=100,
ymin=1e-2,
ymax=2e-1,
xlabel={\textit{v} [\si{km/h}]},
ylabel={MSE},
grid=both,
ymode=log
]
\addplot[line width=1,TUMBeamerRed,solid,mark=x]
table[x=velocity,y=mse] {M16_N4_1Path_Snr0_WienerGenie.txt};
\addlegendentry{LMMSE Perfect};
\addplot[line width=1,solid,mark=diamond]
table[x=velocity,y=mse] {M16_N4_1Path_Snr0_WienerJakes.txt};
\addlegendentry{LMMSE Jakes};
\addplot[line width=1,TUMBeamerOrange,solid,mark=Mercedes star]
table[x=velocity,y=mse] {M16_N4_1Path_Snr0_Gridded.txt};
\addlegendentry{Gridded};
\addplot[line width=1,TUMBeamerGreen,solid,mark=square]
table[x=velocity,y=mse] {M16_N4_1Path_Snr0_StructuredToep.txt};
\addlegendentry{Structured Toep};
\addplot[line width=1,TUMBeamerBlue,solid,mark=triangle]
table[x=velocity,y=mse] {M16_N4_1Path_Snr0_StructuredCirc.txt};
\addlegendentry{Structured Circ};
\addplot[line width=1,TUMBeamerLightBlue,solid,mark=star]
table[x=velocity,y=mse] {M16_N4_1Path_Snr0_NNToep.txt};
\addlegendentry{NN Toep};
\addplot[line width=1,TUMBeamerDarkRed,solid,mark=o]
table[x=velocity,y=mse] {M16_N4_1Path_Snr0_NNCirc.txt};
\addlegendentry{NN Circ};
\end{axis}
\end{tikzpicture}
}

\caption{MSE at prediction step $l=4$, $M=16$, SNR $=0\si{dB}$, $P=1$, $f_c=2\si{GHz}$, $T_{s} = 20.57\si{\micro\second}$, $N=16$ or $N=32$ }
\label{fig:M16_N4_1Path_Snr0}
\end{figure}

In the next simulation setting (Fig. \ref{fig:M16_N4_1Path_Snr0}), the number of observed symbols remains $M = 16$ and the prediction step remains as well unchanged, $l=4$. The SNR is now $0\si{dB}$ and the number of impinging plane-waves at the user is again one, i.e., $P=1$. The number of samples for the construction of the predictors is set to $N=16$ or $N=32$ (depending on $\mathbf{Q}$).

As in the previous simulation, the MSE of the LMMSE Perfect predictor remains constant for all velocities. Obviously, the LMMSE Perfect predictor again outperforms the LMMSE Jakes predictor for all velocities. The Structured Circ predictor, the Gridded predictor, the Structured Toep predictor and the LMMSE Jakes predictor perform equally well. The Neural Network Predictors are again initialized as described above and for training them again 3000 mini-batches, each of size $B=50$, were used. After training the Neural Network Predictors, both of them outperform the LMMSE Jakes predictor and the Gridded predictor for all considered velocities. The NN Toep predictor has a slightly lower MSE as compared to the NN Circ predictor for velocities larger than $60\si{km/h}$ (Fig. \ref{fig:M16_N4_1Path_Snr0}). Both of the Neural Network predictors outperform the LMMSE Perfect predictor for the velocity range from $0\si{km/h}$ to $90\si{km/h}$ (Fig. \ref{fig:M16_N4_1Path_Snr0}).
\\

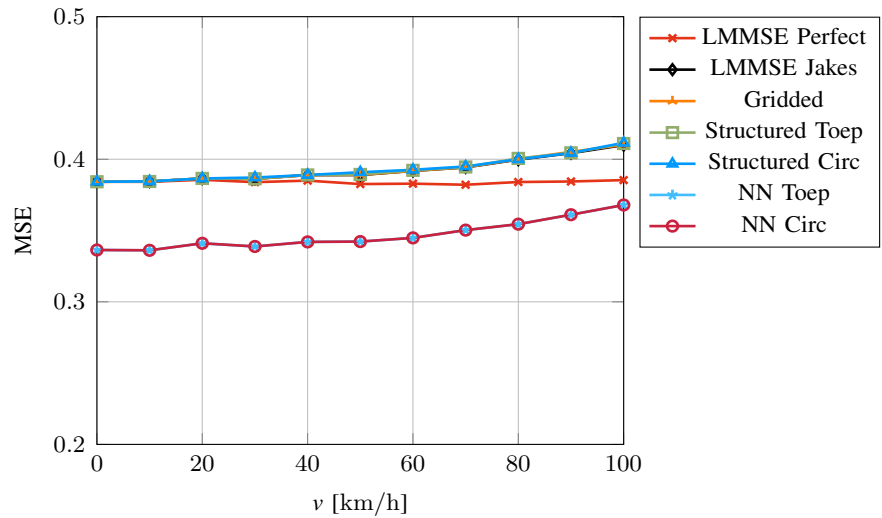
\begin{figure}[H]
    \centering
    \resizebox{340pt}{200pt}{
    \begin{tikzpicture}
\begin{axis}[
legend pos= outer north east,
legend style={font=\footnotesize},
xmin=0,
xmax=100,
ymin=2e-1,
ymax=5e-1,
xlabel={\textit{v} [\si{km/h}]},
ylabel={MSE},
grid=both,
]
\addplot[line width=1,TUMBeamerRed,solid,mark=x]
table[x=velocity,y=mse] {M16_N4_1Path_Snrmin10_WienerGenie.txt};
\addlegendentry{LMMSE Perfect};
\addplot[line width=1,solid,mark=diamond]
table[x=velocity,y=mse] {M16_N4_1Path_Snrmin10_WienerJakes.txt};
\addlegendentry{LMMSE Jakes};
\addplot[line width=1,TUMBeamerOrange,solid,mark=Mercedes star]
table[x=velocity,y=mse] {M16_N4_1Path_Snrmin10_Gridded.txt};
\addlegendentry{Gridded};
\addplot[line width=1,TUMBeamerGreen,solid,mark=square]
table[x=velocity,y=mse] {M16_N4_1Path_Snrmin10_StructuredToep.txt};
\addlegendentry{Structured Toep};
\addplot[line width=1,TUMBeamerBlue,solid,mark=triangle]
table[x=velocity,y=mse] {M16_N4_1Path_Snrmin10_StructuredCirc.txt};
\addlegendentry{Structured Circ};
\addplot[line width=1,TUMBeamerLightBlue,solid,mark=star]
table[x=velocity,y=mse] {M16_N4_1Path_Snrmin10_NNToep.txt};
\addlegendentry{NN Toep};
\addplot[line width=1,TUMBeamerDarkRed,solid,mark=o]
table[x=velocity,y=mse] {M16_N4_1Path_Snrmin10_NNCirc.txt};
\addlegendentry{NN Circ};
\end{axis}
\end{tikzpicture}
}
\caption{MSE at prediction step $l=4$, $M=16$, SNR $=-10\si{dB}$, $P=1$, $f_c=2\si{GHz}$, $T_{s} = 20.57\si{\micro\second}$, $N=16$ or $N=32$ }
\label{fig:M16_N4_1Path_Snrmin10}
\end{figure}

In the next simulation (Fig. \ref{fig:M16_N4_1Path_Snrmin10}), we set the SNR to $-10\si{dB}$. The number of observed symbols remains $M = 16$ and the prediction step remains as well unchanged, $l=4$. The number of impinging plane-waves at the user remains one, i.e., $P=1$. The number of samples for the construction of the predictors is still set to $N=16$ or $N=32$ (depending on $\mathbf{Q}$).

Similar to the previous simulations, the MSE of the LMMSE Perfect predictor remains constant for all velocities and again outperforms the LMMSE Jakes predictor for all velocities. The Structured Circ predictor, the Gridded predictor and Structured Toep predictor perform again almost equally well as the LMMSE Jakes predictor. We train the neural networks with 3000 mini-batches, each of size $B=50$. In this simulation setting, the trained Neural Network Predictors outperform all other predictors for all considered velocities.
\\

\begin{figure}[H]
    \centering
    \resizebox{340pt}{200pt}{
    \begin{tikzpicture}
\begin{axis}[
legend pos= outer north east,
legend style={font=\footnotesize},
xmin=0,
xmax=100,
ymin=5e-2,
ymax=1e-1,
xlabel={\textit{v} [\si{km/h}]},
ylabel={MSE},
grid=both,
scaled y ticks = false,
y tick label style={/pgf/number format/fixed}
]
\addplot[line width=1,TUMBeamerRed,solid,mark=x]
table[x=velocity,y=mse] {M16_N4_2Paths_Snr0_WienerGenie.txt};
\addlegendentry{LMMSE Perfect};
\addplot[line width=1,solid,mark=diamond]
table[x=velocity,y=mse] {M16_N4_2Paths_Snr0_WienerJakes.txt};
\addlegendentry{LMMSE Jakes};
\addplot[line width=1,TUMBeamerOrange,solid,mark=Mercedes star]
table[x=velocity,y=mse] {M16_N4_2Paths_Snr0_Gridded.txt};
\addlegendentry{Gridded};
\addplot[line width=1,TUMBeamerGreen,solid,mark=square]
table[x=velocity,y=mse] {M16_N4_2Paths_Snr0_StructuredToep.txt};
\addlegendentry{Structured Toep};
\addplot[line width=1,TUMBeamerBlue,solid,mark=triangle]
table[x=velocity,y=mse] {M16_N4_2Paths_Snr0_StructuredCirc.txt};
\addlegendentry{Structured Circ};
\addplot[line width=1,TUMBeamerLightBlue,solid,mark=star]
table[x=velocity,y=mse] {M16_N4_2Paths_Snr0_NNToep.txt};
\addlegendentry{NN Toep};
\addplot[line width=1,TUMBeamerDarkRed,solid,mark=o]
table[x=velocity,y=mse] {M16_N4_2Paths_Snr0_NNCirc.txt};
\addlegendentry{NN Circ};
\end{axis}
\end{tikzpicture}
}
\caption{MSE at prediction step $l=4$, $M=16$, SNR $=0\si{dB}$, $P=2$, $f_c=2\si{GHz}$, $T_{s} = 20.57\si{\micro\second}$, $N=16$ or $N=32$ }
\label{fig:M16_N4_2Paths_Snr0}
\end{figure}
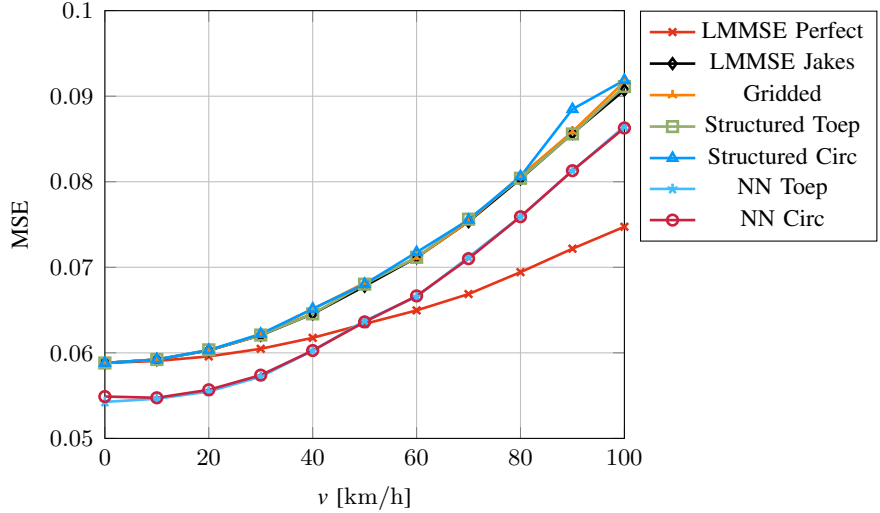

So far we have only considered the case of only one impinging plane-wave at the user. In the next simulation (Fig. \ref{fig:M16_N4_2Paths_Snr0}), we now increase the number of impinging plane-waves at the user to two, i.e., $P=2$. The number of observed symbols remains $M = 16$ and the prediction step remains as well unchanged, $l=4$. The SNR is $0\si{dB}$. The number of samples for the construction of the predictors is still set to $N=16$ or $N=32$ (depending on $\mathbf{Q}$).

In contrast to the previous simulations, the MSE of the LMMSE Perfect predictor does not remain constant for all velocities due to the increased number of paths to two. The LMMSE Jakes predictor with $P\to\infty$ suggests that the predictor performance heavily depends on the Doppler bandwidth $B_D$ of the system setup (see \eqref{coorelationFunc_infP}), which depends on the velocity of the user. 
The Structured Circ predictor, the Gridded predictor, the Structured Toep predictor and the LMMSE Jakes predictor perform again almost equally well. We train the neural networks with 3000 mini-batches, each of size $B=50$, where each channel realization is constructed with $P=2$. In this simulation setting, the trained Neural Network Predictors outperform all other predictors for the velocity range from $0\si{km/h}$ to $50\si{km/h}$ (Fig. \ref{fig:M16_N4_2Paths_Snr0}).
\\

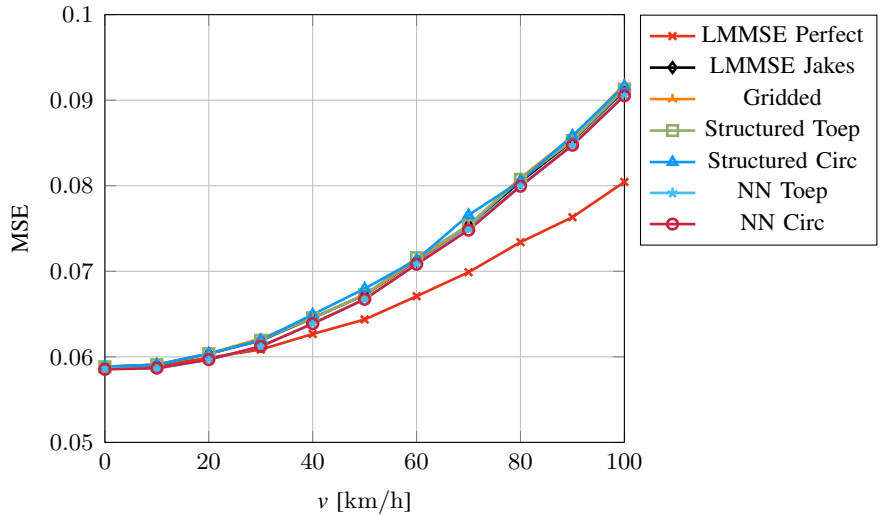
\begin{figure}[H]
    \centering
    \resizebox{340pt}{200pt}{
    \begin{tikzpicture}
\begin{axis}[
legend pos= outer north east,
legend style={font=\footnotesize},
xmin=0,
xmax=100,
ymin=5e-2,
ymax=1e-1,
xlabel={\textit{v} [\si{km/h}]},
ylabel={MSE},
grid=both,
scaled y ticks = false,
y tick label style={/pgf/number format/fixed}
]
\addplot[line width=1,TUMBeamerRed,solid,mark=x]
table[x=velocity,y=mse] {M16_N4_3Paths_Snr0_WienerGenie.txt};
\addlegendentry{LMMSE Perfect};
\addplot[line width=1,solid,mark=diamond]
table[x=velocity,y=mse] {M16_N4_3Paths_Snr0_WienerJakes.txt};
\addlegendentry{LMMSE Jakes};
\addplot[line width=1,TUMBeamerOrange,solid,mark=Mercedes star]
table[x=velocity,y=mse] {M16_N4_3Paths_Snr0_Gridded.txt};
\addlegendentry{Gridded};
\addplot[line width=1,TUMBeamerGreen,solid,mark=square]
table[x=velocity,y=mse] {M16_N4_3Paths_Snr0_StructuredToep.txt};
\addlegendentry{Structured Toep};
\addplot[line width=1,TUMBeamerBlue,solid,mark=triangle]
table[x=velocity,y=mse] {M16_N4_3Paths_Snr0_StructuredCirc.txt};
\addlegendentry{Structured Circ};
\addplot[line width=1,TUMBeamerLightBlue,solid,mark=star]
table[x=velocity,y=mse] {M16_N4_3Paths_Snr0_NNToep.txt};
\addlegendentry{NN Toep};
\addplot[line width=1,TUMBeamerDarkRed,solid,mark=o]
table[x=velocity,y=mse] {M16_N4_3Paths_Snr0_NNCirc.txt};
\addlegendentry{NN Circ};
\end{axis}
\end{tikzpicture}
}
\caption{MSE at prediction step $l=4$, $M=16$, SNR $=0\si{dB}$, $P=3$, $f_c=2\si{GHz}$, $T_{s} = 20.57\si{\micro\second}$, $N=16$ or $N=32$ }
\label{fig:M16_N4_3Paths_Snr0}
\end{figure}

In the next simulation (Fig. \ref{fig:M16_N4_3Paths_Snr0}), we now further increase the number of impinging plane-waves at the user to three, i.e., $P=3$. The number of observed symbols remains $M = 16$ and the prediction step remains as well unchanged, $l=4$. The SNR is as in the previous simulation again set to $0\si{dB}$. The number of samples for the construction of the predictors is still set to $N=16$ or $N=32$ (depending on $\mathbf{Q}$).

We see that the LMMSE Perfect predictor outperforms the LMMSE Jakes predictor. However, as compared to the previous simulation setting with $P=2$, a smaller gap is present between the LMMSE Perfect and the LMMSE Jakes predictor (due to the fact that $P=3$). The Structured Circ predictor, the Gridded predictor, the Structured Toep predictor and the LMMSE Jakes predictor perform again almost equally well. We again train the neural networks with 3000 mini-batches, each of size $B= 50$, where each channel realization is constructed now with $P=3$. The Neural Network Predictors are slightly performing better than the LMMSE Perfect predictor for velocities up to $25\si{km/h}$. For all other velocities the neural network based predictors are outperforming all other predictors, except the LMMSE Perfect predictor. The simulation settings from Fig. \ref{fig:M16_N4_1Path_Snr0}, Fig. \ref{fig:M16_N4_2Paths_Snr0} and Fig. \ref{fig:M16_N4_3Paths_Snr0}, only differ in their number of paths $P$.

We considered simulation settings with low path numbers, since with a relatively high number of propagation paths the channel is similar to Jakes model.

\section{Conclusion}
A novel approach to learn a feed-forward neural network channel predictor was presented in this paper. Starting from the LMMSE predictor a reformulated version was derived. By making two key assumption it was possible to further derive predictors, which are motivated by the structure of the MMSE predictor. The Neural Network Predictor is initialized with the Structured Predictor. By further training the network based predictors, it is possible to compensate the approximation errors due to the assumptions we made. Simulation results show that the Neural Network Predictors outperform the LMMSE predictor based on the assumption of a Jakes spectrum. For the specific cannel model assumptions in this paper, the Neural Network predictor even outperforms the LMMSE predictor based on the known covariance function for a low number of paths $P$ and low SNR values.


\ifCLASSOPTIONcaptionsoff
  \newpage
\fi

\bibliographystyle{IEEEtran}
\vspace{10pt}
\bibliography{Turan_Utschick_Prediction_2019}

\end{document}